\documentclass[]{aastex631}

\usepackage{multirow}
\usepackage{booktabs, makecell}
\usepackage{amssymb}
\usepackage{xcolor}
\def\r{\textcolor{red}}
\def\b{\textcolor{blue}}
\def\p{\textcolor{purple}}
\def\w{\textcolor{brown}}
\def\g{\textcolor{green}}

\usepackage{ulem}

\begin{document}

\title{Revisiting the flaring activity in early 2015 of BL Lacertae object S5 0716+714}

\author[0009-0000-4102-9115]{Zhihao Ouyang}
\affiliation{Shanghai Key Lab for Astrophysics, Shanghai Normal University, Shanghai, 200234, People's Republic of China}
\affiliation{Center for Astrophysics, Guangzhou University, Guangzhou, 510006, People's Republic of China}

\author[0000-0001-8244-1229]{Hubing Xiao}
\affiliation{Shanghai Key Lab for Astrophysics, Shanghai Normal University, Shanghai, 200234, People's Republic of China}

\author[0000-0003-1530-3031]{Marina Manganaro}
\affiliation{Department of Physics, University of Rijeka, Rijeka, 51000, Croatia}

\author{Shangchun Xie}
\affiliation{Shanghai Key Lab for Astrophysics, Shanghai Normal University, Shanghai, 200234, People's Republic of China}

\author{Jingyu Wu}
\affiliation{Shanghai Key Lab for Astrophysics, Shanghai Normal University, Shanghai, 200234, People's Republic of China}

\author{Jianzhen Chen}
\affiliation{Shanghai Key Lab for Astrophysics, Shanghai Normal University, Shanghai, 200234, People's Republic of China}

\author{Rui Xue}
\affiliation{Department of Physics, Zhejiang Normal University, Jinhua, 321004, People's Republic of China}

\author{Gege Wang}
\affiliation{Key Laboratory of Cosmology and Astrophysics (Liaoning) $\&$ College of Sciences, Northeastern University, Shenyang, 110819, People's Republic of China}

\author{Shaohua Zhang}
\affiliation{Shanghai Key Lab for Astrophysics, Shanghai Normal University, Shanghai, 200234, People's Republic of China}

\author[0000-0002-5929-0968]{Junhui Fan}
\affiliation{Center for Astrophysics, Guangzhou University, Guangzhou, 510006, People's Republic of China}
\affiliation{Great Bay Brand Center of the National Astronomical Data Center, Guangzhou, 510006, People's Republic of China}
\affiliation{Key Laboratory for Astronomical Observation and Technology of Guangzhou, Guangzhou, 510006, People's Republic of China}
\affiliation{Astronomy Science and Technology Research Laboratory of Department of Education of Guangdong Province, Guangzhou, 510006, People's Republic of China}

\correspondingauthor{Hubing Xiao, Marina Manganaro, Jianzhen Chen}
\email{hubing.xiao@shnu.edu.cn, marina.manganaro@phy.uniri.hr, jzchen@shnu.edu.cn}



\begin{abstract}

In this work, we analyzed multi-wavelength data of the BL Lac object S5 0716+714 to investigate its emission mechanisms during a flaring state observed in early 2015. 
We examined the temporal behavior and broadband spectral energy distributions (SEDs) during the flare. 
The size of the $\gamma$-ray emission region was estimated based on the variability timescale. 
To explore the multi-wavelength properties of S5 0716+714, we employed three one-zone models: the SSC model, the SSC plus EC model, and the SSC plus \textit{pp} interactions model, to reproduce the SEDs.
Our findings indicate that while the SSC model can describe the SEDs, it requires an extreme Doppler factor. 
In contrast, the SSC plus EC model successfully fits the SEDs under the assumption of weak external photon fields but requires a high Doppler factor. 
Additionally, the SSC plus \textit{pp} interactions model also reproduces the SEDs, with $\gamma$-ray emission originating from $\pi^{0}$ decay. 
However, this model leads to a jet power that exceeds the Eddington luminosity, which remains plausible due to the flaring state or the presence of a highly collimated jet.

\end{abstract}



\section{Introduction} \label{Intro}

Blazars, a unique subclass of active galactic nuclei (AGNs), exhibit extreme observational characteristics including strong variability, high polarization, superluminal motion, and $\gamma$-ray radiation \citep[e.g.,][and references therein]{Urry1995, Fan2004, Lyutikov2017, Lister2019ApJ, Xiao2019SCPMA, Xiao2022MNRAS, Abdollahi2022}.
Blazars are grouped into two subclasses based on their optical continuum: BL Lacertae objects (BL Lacs), which show weak or absent emission lines (equivalent width, EW $<$ 5 $\rm \mathring{A}$), and flat spectrum radio quasars (FSRQs), which show prominent emission lines \citep[EW $\geqslant$ 5 $\rm \mathring{A}$;][]{Stickel1991, Urry1995, Scarpa1997}.
The spectral energy distribution (SED) exhibits a distinct two-hump structure. 
The low-energy hump, observed in the infrared to X-ray range, is attributed to a synchrotron emission of relativistic electrons.
The high-energy hump, located at MeV to GeV energies, can be generated either through the inverse Compton process (IC) in a leptonic scenario \citep{Tavecchio1998ApJ, Ghisellini2009MNRAS, Tan2020ApJS}, or through a hadronic model \citep[e.g.,][]{Mucke2003, Bottcher2009ApJ, Cerruti2015MNRAS, Cerruti2019MNRAS, Gao2019NatAs3, Xue2022PhRvD106}. 

S5 0716+714 (4FGL J0721.9+7120) is classified as an intermediate-peak-frequency BL Lac (IBL) object according to \citet{Fan2016APJS}, located at a distance of $z=0.2304\pm0.0013$ \citep{Pichel2023A&A}.  
It was first discovered in the late 1970s in the radio band \citep{Perley1980AJ, Kuehr1981A&AS}, and early radio observations showed strong emission and notable variability \citep{Kraus2003A&A}. 
Subsequent high-resolution very long baseline interferometry (VLBI) observations showed that S5 0716+714's jet components exhibit fast superluminal motion, indicating it has a strong Doppler beaming effect \citep[e.g.,][]{Bach2005A&A, Rani2015A&A}.
Optical observations have been pivotal in studying the variability of S5 0716+714.
Previous studies demonstrated rapid variability on timescales of hours to days \citep{Wagner1996AJ, Poon2009ApJS, Gupta2012MNRAS, Tripathi2024MNRAS}, with intraday variability suggesting a compact emission region for S5 0716+714.
Long-term optical monitoring programs have reported a bluer-when-brighter behavior \citep[e.g.,][]{Dai2015ApJS, Xiong2020ApJS}.
Additionally, \citet{Ikejiri2011PASJ} performed the photopolarimetric monitoring observations, revealing complex variability in both the degree and angle of polarization, supporting the hypothesis that the optical emission primarily originates from synchrotron radiation. 
The Roentgen Satellite (ROSAT) provided the first X-ray detection and it showed significant rapid variability and a double power-law fitted spectrum, implying a mixture of synchrotron and inverse Compton components in the X-ray band \citep{Cappi1994MNRAS}.
More recent X-ray observations have offered detailed spectral and temporal analyses, they showed that the X-ray fluxes were highly variable and the break energy between the synchrotron and inverse Compton components shifted during different flux states \citep{Foschini2006A&A, Wierzcholska2015MNRAS, Wierzcholska2016MNRAS}.
The source was first detected in the $\gamma$-ray band by the Energetic Gamma Ray Experiment Telescope (EGRET) on board the Compton Gamma-Ray Observatory (CGRO) and has since been detected several times at different flux levels \citep{Lin1995ApJ, Hartman1999ApJS}. 
It is one of the brightest and most variable sources in the $\gamma$-ray band, with a variability index (VI) of 3680.86 reported by the \textit{Fermi} Large Area Telescope \citep[\textit{Fermi}-LAT;][]{Abdollahi2022}. 
Several studies have attempted to explore the $\gamma$-ray activity of S5 0716+714.
For example, \citet{Rani2014A&A} reported a significant correlation between $\gamma$-ray fluxes and position angle variations in the VLBI jet, while \citet{Geng2020ApJ} found a highly variable $\gamma$-ray flux with a spectral break between 0.93 and 6.90 GeV through long-term observations. 
Simultaneously multi-wavelength observation has been employed as an effective method to study the blazar emission mechanisms. 
\citet{Rani2013A&A552} conducted a comprehensive campaign, including radio, optical, X-ray, and $\gamma$-ray observation, to investigate a detailed temporal behavior and constructed a broadband detailed SED.
Similarly, \citet{Liao2014ApJ} performed a multi-wavelength study of S5 0716+714, finding significant variability and correlation across all bands, and suggesting that the synchrotron self-Compton (SSC) plus external Compton (EC) model is preferred to describe the broadband SED.

The first very-high-energy (VHE) $\gamma$-ray detection of S5 0716+714 was performed by Major Atmospheric Gamma-ray Imaging Cherenkov telescopes (MAGIC) in 2007 \citep{Anderhub2009ApJ}.  In the latter work, the observation gave a significance of 5.8$\sigma$ over 13.1 hours of observation in November 2007, and a significance of 6.9$\sigma$ in April 2008. 
Interestingly, the VHE observation coincided with optical high-state emission, implying a possible correlation between VHE and optical emission. 
This led to the exploration of the one-zone SSC model and the structured (``spine+layer'') jet model \citep{Anderhub2009ApJ}.
In late December 2014, S5 0716+714 became brighter in the optical and infrared bands, exhibiting an exceptionally high state in January 2015, with the highest flux recorded in these bands \citep{Carrasco2015ATel6902, Bachev2015ATel6944, Bachev2015ATel6957}. 
MAGIC observations triggered by this flare revealed a potentially variable VHE flux ranging from 4$\times$10$^{-11}$ cm$^{-2}$ s$^{-1}$ to 7$\times$10$^{-11}$ cm$^{-2}$ s$^{-1}$ above 150 GeV between January 22 and January 26, 2015 \citep{Mirzoyan2015ATel6999, Marina2018}. 
This activity was studied in detail in \citet{Marina2018} based on the multi-wavelength observations of its flaring behavior in Jan 2015. 
They stated that, due to the high level of optical flux, the broadband SED of the source can not be reproduced by a one-zone SSC model.
Instead, an interaction between a superluminal knot and a recollimation knot was found, implying a two-zone model was preferred. 
However, the model underestimates the $\gamma$-ray flux in 10$\sim$100 GeV. 
In addition, the electric vector position angle (EVPA) showed a fast rotation of $\sim$360$^{\circ}$ and the high energy $\gamma$-ray flare occurring during the $\gamma$-ray flaring activity, suggesting a shock-shock interaction in the jet. 
Considering the polarization variations during the flare, \citet{Chandra2015ApJ} suggested that the magnetic reconnections were likely involved in this flare. 
There is no doubt that it is necessary to further study the radiation mechanism of the S5 0716+714 flare that occurred in Jan 2015. 

In this work, we aim to further explore the high-energy emission of S5 0716+714 during the January and February 2015 flare. 
We will report on the multi-wavelength campaign involving \textit{Swift}, \textit{NuSTAR}, \textit{Fermi}, and MAGIC observations, and investigate the temporal behavior and spectral properties of the source. 
Additionally, we aim to restructure the emission region and reproduce the broadband SED with a new hybrid model.

The paper is structured as follows.
The observation introduction and the data reduction are presented in Section \ref{Obs}.  
The result and discussion are presented in Section \ref{result_discussion}.
Finally, the summary is given in Section \ref{Summary}.

\section{Observation and data reduction} \label{Obs}

\subsection{Swift observation}

The Neil Gehrels \textit{Swift} Gamma-Ray Burst Observatory \citep[\textit{Swift};][]{Gehrels2004} was launched in 2004 and includes three instruments: the Ultraviolet and Optical Telescope \citep[UVOT;][]{Roming2005}, the X-ray telescope \citep[XRT;][]{Burrows2004} and the Burst Alert Telescope \citep[BAT;][]{Barthelmy2005}. 
The analysis was performed using the HEASoft package (v6.31.1) released by the NASA High Energy Astrophysics Archive Research Center (HEASARC).

\subsubsection{Swift-UVOT}

The \textit{Swift}-UVOT observations \citep{Roming2005} include three optical ($v$, $b$, and $u$) and three UV ($w$1, $m$2, and $w$2) photometric bands \citep{Poole2008, Breeveld2010}. 
The source region was extracted from a circular region of 5$^{\prime \prime}$ centered on the source, and the background region was extracted from a circular region of 20$^{\prime \prime}$ near the source for all filters.
The \texttt{uvotmaghist} task was used to analyze all filter data and produce the photometric light curve data using the calibration from the release of CALDB (version 20211108). 
All the UVOT data were checked for the small-scale sensitivity inhomogeneities, which occur when the source falls within the small areas of low sensitivity\footnote{\url{https://swift.gsfc.nasa.gov/analysis/uvot_digest/sss_check.html}}. 
In addition, photometric data in which the source was saturated were excluded. 
Galactic extinction was corrected for the observed magnitude with a value of $E(B-V) = 0.0268$ \citep{Schlafly2011} using the interstellar extinction law with $R_{V} = 3.1$ \citep{Fitzpatrick1999}. 
Finally, the corrected magnitudes were converted into fluxes using the standard zero points from \citet{Breeveld2011}.

\subsubsection{Swift-XRT}

For S5 0716+714, the \textit{Swift}-XRT operated in Photon Counting (PC) and Windowed Timing (WT) readout modes with a total exposure time of $\sim$ 1.55$ \times $10$^{5}$ s. 
The data were processed using the XRTDAS software package (v3.7.0) with the release of CALDB (version 20220803). 
The cleaned events were produced using the \texttt{xrtpipeline} task, selecting events with grades 0$-$12 for PC mode and grades 0$-$2 for WT mode. 
For the PC mode, the source region was extracted from a circular region of 20 pixels ($\sim$ 47$^{\prime \prime}$) centered on the source.
If the count rate was above 0.5 $\rm count \cdot s^{-1}$, pile-up correction was applied by excluding the central region events within 3$^{\prime \prime}$$-$10$^{\prime \prime}$, using an annulus with an outer radius of 20 pixels. 
The background region was extracted from an annulus with an inner radius of 80$^{\prime \prime}$ and an outer radius of 160$^{\prime \prime}$ centered on the source. 
For the WT mode, the source and background regions were extracted from circular regions of 20$-$30 pixels, depending on source brightness and exposure time, with the source region centered on the source and the background region nearby. 
The high-level product data, including the spectra and ancillary response files (ARF), were generated from the cleaned data using the \texttt{xrtproducts} task. 
The spectra were grouped using the \texttt{grppha} (v3.1.0) tool to ensure at least one count per bin for fitting with Cash statistic \citep{Cash1979}. 
The grouped spectra were loaded into \texttt{XSPEC} (v12.13.0c) and fitted with an absorbed power-law model with normalization energy $E_{0}$ = 1 keV. 
The Galactic hydrogen column density was fixed at $n_{\rm H}$ = 2.88$ \times $10$^{20}$ $\rm cm^{-2}$ \citep{HI4PI2016}. 
In \texttt{XSPEC} settings, the solar abundances used in the photoelectric absorption models were set as \texttt{wilm} \citep{Wilms2000} and the photoionization absorption cross-sections were set as \texttt{vern} \citep{Verner1996}. 
For X-ray data fitting, the parameter errors correspond to 90\% confidence errors ($\Delta \chi^{2} = 2.706$). 
Finally, the unabsorbed fluxes and photon spectral indices were obtained.

\subsection{NuSTAR observation}

The Nuclear Spectroscopic Telescope Array (\textit{NuSTAR}), launched in 2012, operates in the hard X-ray range (3$-$79 keV) and features two telescopes with multilayer coatings that focus reflected X-rays onto pixillated CdZnTe focal plane modules, FPMA and FPMB. 
The observation provides a spectral resolution of approximately 1 keV, and the half-power diameter of an image of a point source is $\sim$1$^{\prime}$. 
Additional information can be found in \cite{Harrison2013}.

\textit{NuSTAR} observed S5 0716+714 with its two focal plane modules on 24 January 2015 (MJD 57046), with an exposure time of $\sim$ 18.5 ks. 
The raw data were processed with the \textit{NuSTAR} Data Analysis Software (NuSTARDAS, v2.1.2) package using the calibration from the release of CALDB (version 20230307). 
The cleaned event files were produced by the \texttt{nupipeline} task. 
The source region was extracted from a circular region of 45$^{\prime \prime}$ centered on the centroid of X-ray emission. 
The background was extracted from a position 5$^{\prime}$ away from the centroid of the X-ray emission, using a circular region of 1.5$^{\prime}$. 
The spectra were produced from the cleaned event files and grouped with at least one count per bin using the \texttt{nuproducts} task. 
We focused on the energy range of 3$-$60 keV where the source was detected. 
The \texttt{XSPEC} settings (including the Galactic hydrogen column density) were the same as those used in the \textit{Swift}-XRT analysis.

\subsection{Fermi-LAT observation}

The \textit{Fermi} Gamma-ray Space Telescope was launched in 2008 and it consists of two instruments: Gamma-ray Burst Monitor (GBM) and Large Area Telescope (LAT). 
The \textit{Fermi}-LAT \citep{Atwood2009} is capable of detecting $\gamma$-ray in the energy range from 20 MeV to beyond 300 GeV. 
The point source sensitivity of \textit{Fermi}-LAT is $\sim$ 2$\times$10$^{-13}$ erg cm$^{-2}$ s$^{-1}$ for the north Celestial pole after 10 yrs operation \citep{Ajello2021ApJS256}.

We used data events from the \textit{Fermi}-LAT's Pass 8 database in the period from MJD 57010 (2014 Dec 19) to 57075 (2015 Feb 22). 
The data were collected within an energy range of 0.1$-$100 GeV and within a 15$^{\circ}$ radius region of interest (ROI) centered on S5 0716+714. 
A maximum zenith angle value of 90$^{\circ}$ was selected to avoid background $\gamma$-rays from the Earth's limb. 
We performed an unbinned likelihood analysis of the data using the latest \texttt{Fermitools} \citep[v2.2.0;][]{Fermitools2019ascl}
and the instrument response functions (IRFs) \texttt{P8R3\_SOURCE\_V3}. 
The conditions ``\texttt{evclass=128, evtype=3}" were used to filter events with a high probability of being photons and ``\texttt{(DATA\_QUAL$\geqslant$0)\&\&(LAT\_CONFIG==1)}" was used to select the good time intervals. 
The model file, generated by \texttt{make4FGLxml} python package, included all the sources from the Fermi-LAT Fourth Source catalog \citep[4FGL-DR4;][]{Abdollahi2022} within 25$^{\circ}$ of S5 0716+714 as well as the Galactic (\texttt{gll\_iem\_v07.fits}) and extragalactic isotropic (\texttt{iso\_P8R3\_SOURCE\_V3\_v1.txt}) diffuse emission components. 
The spectral parameters of sources with an average significance larger than 5$\sigma$ within 5$^{\circ}$ of the ROI were left free, as well as sources within 10$^{\circ}$ of the ROI with a variable index $\geqslant$ 24.725 \citep{Abdollahi2022}.
The best model between a power-law
$\left[ {\rm PL; \,}
    \frac{\mathrm{d} N}{\mathrm{d} E} = N_{0} \left( \frac{E}{E_{0}}\right) ^{-\Gamma_{\gamma}}
\right]$
model and a log-parabola
$\left[ {\rm LP; \,}
    \frac{\mathrm{d} N}{\mathrm{d} E} = N_{0} \left( \frac{E}{E_{0}}\right) ^{-\left( \alpha + \beta \log \left( \frac{E}{E_{0}}\right) \right)}
\right]$\footnote{Here, the `$\log$' refers to the decimal logarithm, whereas `$\ln$' in Eq. (\ref{BIC}) below denotes the natural logarithm.} model was selected by calculating $\rm TS_{curve} =  2(\log \mathcal{L}_{LP} - \log \mathcal{L}_{PL})$, where $\rm \mathcal{L}_{PL}$/$\rm \mathcal{L}_{LP}$ represent the maximum likelihood value of power-law and log-parabola, respectively \citep{Nolan2012ApJS, Abdollahi2022}. 
If $\rm TS_{curve} \geqslant $ 16, corresponding to 4$\sigma$, the model was switched to the LP model. 
We found that S5 0716+714 preferred the PL model rather than the LP model.  
We generated the light curves binned in the one-day bin. 
In each time bin the normalization parameters of the sources within 5$^{\circ}$ of ROI and the spectral index of the S5 0716+714 were allowed to vary freely during the spectral fitting. 
The rest of the parameters and other source models were frozen. 
The normalization of the two diffuse emission components was also set free in the analysis. 
We only included flux data points that are significantly detected with TS $\geqslant$ 16. 
Meanwhile, we calculated the 95\% confidence level upper limit flux value for the case of TS $<$ 16 using the \texttt{UpperLimits}{\footnote{\url{https://fermi.gsfc.nasa.gov/ssc/data/analysis/scitools/upper_limits.html}}} tool.
While examining the spectral energy distribution, we fixed the spectral indices as the constant value equal to the value fitting over the whole energy range.

\subsection{MAGIC observation}

We compiled the VHE $\gamma$-ray data, including the light curve and spectra, from \citet{Marina2018}, who divided the observation into two periods: Phase A (MJD 57040$-$57050) and Phase B (MJD 57065$-$57070). 
The spectra have been corrected by the extragalactic background light (EBL) absorption using the redshift $z$=0.2304 \citep{Pichel2023A&A} and the EBL model from \citet{Dominguez2011MNRAS}.

\begin{figure}
    \centering
    \includegraphics[width=6 in]{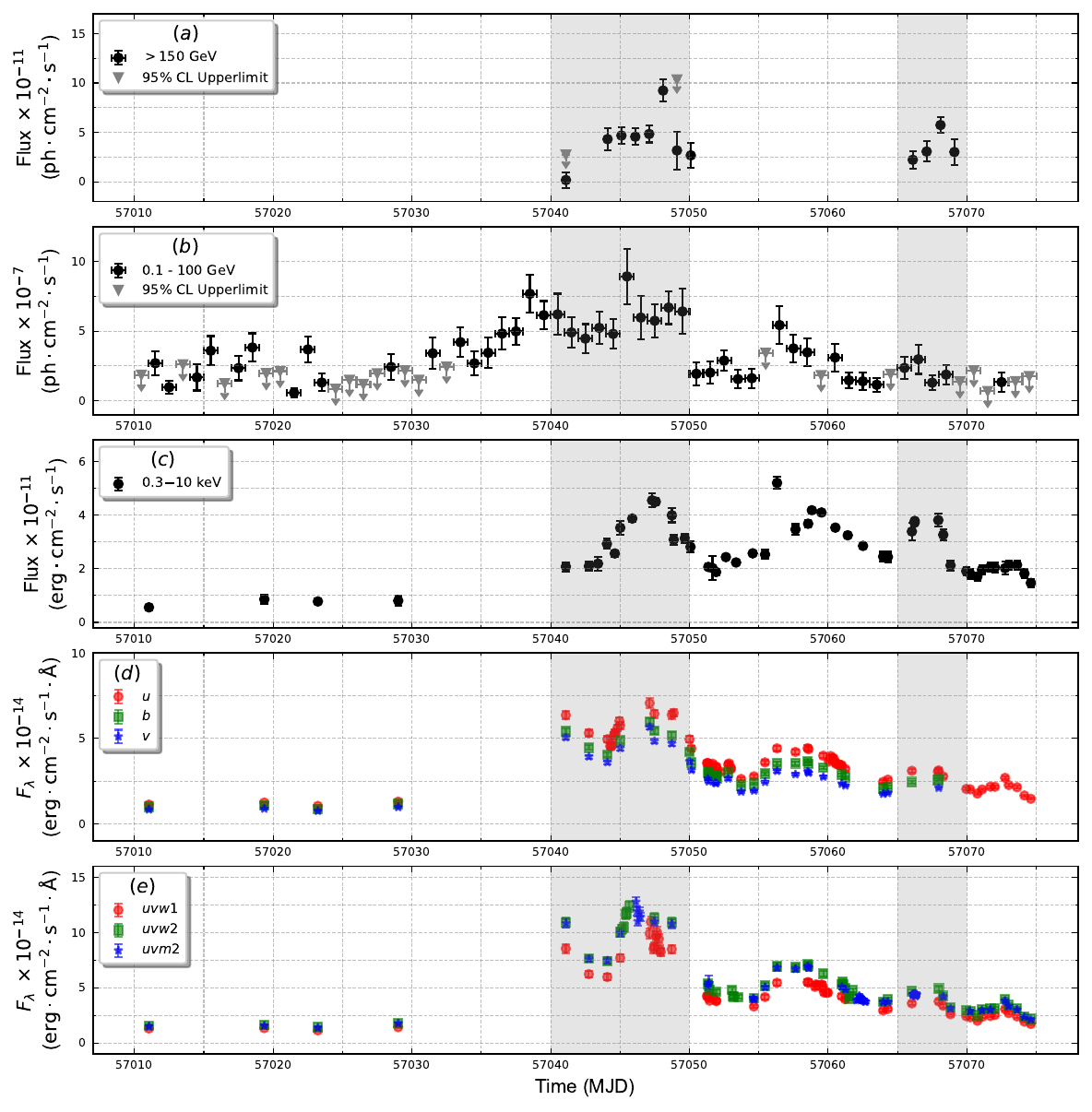}
    \caption{
    The multi-wavelength light curve during the period from MJD 57010 to 57075 with \textit{Swift}, \textit{Fermi}, and MAGIC observations.
     From the top to bottom panels: 
     ($a$) MAGIC VHE flux, $>$ 150 GeV and the gray triangle is the 95\% confidence level upper limit; 
     ($b$) \textit{Fermi}-LAT flux in 0.1$-$100 GeV in one-day bin; 
     ($c$) \textit{Swift}-XRT flux, 0.3$-$10 keV; 
    ($d$) \textit{Swift}-UVOT, $u$, $b$ and $v$ bands with the Galactic extinction correction; 
     ($e$) \textit{Swift}-UVOT, $uvw1$, $uvm2$ and $uvw2$ bands with the Galactic extinction correction. 
     The two gray time intervals are the periods MAGIC observation was taken, meaning Phase A and Phase B.   
     }
    \label{fig:LC}
\end{figure}

\section{Result and Discussion}\label{result_discussion}

\subsection{The $\gamma$-ray variability timescale}

\begin{figure}
    \centering
    \includegraphics[width=5 in]{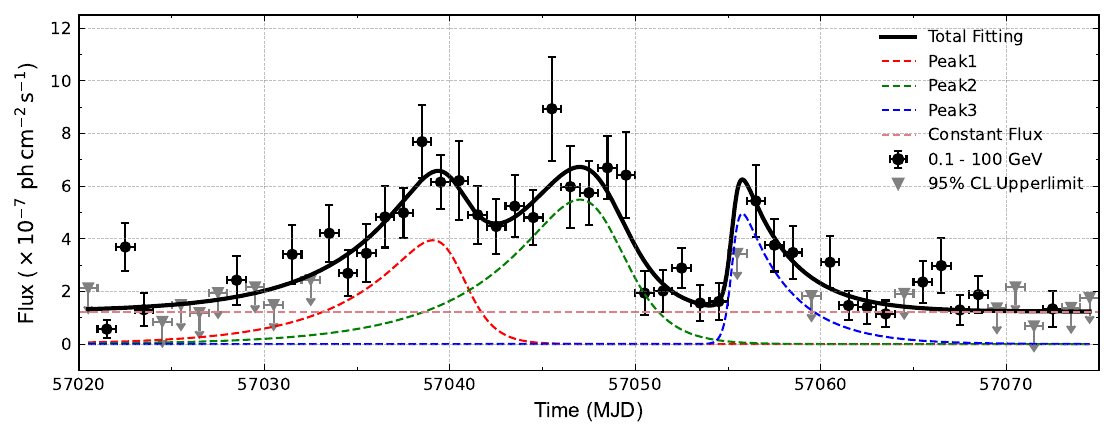}
    \caption{The fitted light curve of the period MJD 57020$-$57075. 
    The dash lines with different colors represent the different peak components and the solid black line represents the sum-fitted time profile of each component. 
    }
    \label{fig:LC_profile}
\end{figure}

For the purpose of studying the geometry of the emission region and the property of the particle population, we modeled the light curve to explore the time profile and variability timescale. 
In this work, we used the \textit{Fermi} $\gamma$-ray data to pursue this task as it is more continuously and uniformly sampled compared to the data of optical, UV, X-ray, and VHE bands. 
The exponential fitting is applied to each component of the flare, the entire light curve is, thus, expressed as a sum of exponential functions with a smoothed transition from raising to falling edge
\begin{equation}
  F(t) = F_{\rm c} + \sum_{i} 2 F_{0, i} \left[  \exp(\frac{t_{0, i}-t}{T_{{\rm r}, i}}) + \exp(\frac{t-t_{0, i}}{T_{{\rm d}, i}})
  \right]^{-1} {\rm ,}
\end{equation}
where $F_{\rm c}$ is the baseline or constant flux, $F_{0}$ is the peak flux value at time $t_{0}$, $T_{\rm r}$ and $T_{\rm d}$ are the rise and decay time, respectively \citep{Abdo2010ApJ722}. 
We constructed a likelihood function for fitting the exponential functions and considered the contribution of the upper limit value to the likelihood function. 
The \texttt{iminuit} package was employed to perform the maximum likelihood fitting and the Bayesian Information Criterion (BIC) was utilized to ascertain the optimal number of exponential functions required to fit the light curve:
\begin{equation}
  {\rm BIC} = k \ln (N) - 2 \ln (\hat \mathcal{L}) {\rm ,}
  \label{BIC}
\end{equation}
where $k$ is the number of model parameters, $N$ is the total number of data points, and $\hat \mathcal{L}$ represents the maximized value of the likelihood function for fitting the exponential functions \citep{Edelson1988ApJ, Wit2012SN66}.
We tested different numbers of exponential functions and selected the model that minimized the BIC value. 
Finally, we used three exponential components to fit the light curve from MJD 57020 to 57075, as shown in Fig. \ref{fig:LC_profile}. 
It is important to note that a shorter time bin light curve can reveal more detailed structures and accurate variability timescales, but this comes with increased flux errors and reduced TS values, which can decrease the fitting quality.

Three peak profiles exhibited asymmetric shapes, reflecting the underlying particle acceleration and cooling mechanisms. 
The first two peaks showed a relatively gradual rise followed by a sharp decay, suggesting either a gradual acceleration of particles or rapid cooling or escape of injected/accelerated electrons, resulting in a radiative cooling timescale shorter than the acceleration timescale \citep{Roy2019MNRAS}.
In contrast, the third minor peak, which was also observed in other bands (see Fig. \ref{fig:LC}) but lacked coverage in the VHE band, exhibited a relatively shorter rise timescale compared to its decay timescale, indicating that the asymmetry may stem from changes in the bulk Lorentz factor, the structure of the emission region's shells \citep{Roy2019MNRAS}, or particle injection processes \citep{Wang2022PASP}. 
Our findings on asymmetry differ from those of \citet{Geng2020ApJ}, who reported symmetric flare profiles. This discrepancy arises because their analysis used shorter binning to construct the light curve, increasing the associated errors and leading to divergent results.

In addition, the rise and decay timescales can serve as a tool for constraining the geometry of the emission region, which will be explored in the following subsection.

\begin{deluxetable}{ccccc}
\tabletypesize{\small}
\tablewidth{5pt} 
\tablecaption{Fitting parameters of light curve in Fig. \ref{fig:LC_profile}. \label{tab:LC_params}}
\tablehead{ 
    Component & $t_{0}$    & $F_{0}$    & $T_{\rm r}$    & $T_{\rm d}$  \\
       & (MJD) & ($\times$ 10$^{-7}$ ph cm$^{-2}$ s$^{-1}$) & (day) & (day)  
}
\colnumbers 
\startdata
    1     & 57040.27$\pm$0.98 & 3.16$\pm$1.38 & 4.34$\pm$2.17  & 0.95$\pm$0.76 \\
    2     & 57048.33$\pm$1.99 & 4.75$\pm$1.76 & 4.75$\pm$3.89  & 1.48$\pm$1.02 \\
    3     & 57055.26$\pm$1.58 & 3.29$\pm$2.39 & 0.23$\pm$0.51  & 2.60$\pm$1.23 \\
\enddata
\tablecomments{
    The constant flux is given as $F_{\rm c}$ = (1.23 $\pm$ 0.28) $\times$ 10$^{-7}$ ph cm$^{-2}$ s$^{-1}$.
}
\end{deluxetable}

\subsection{Modeling spectral energy distributions} \label{SED}

The emission mechanisms of blazars can be better understood by modeling their SEDs. 
In this context, to further investigate the physical origin of the flaring activity of S5 0716+714, we modeled the simultaneous multi-wavelength SEDs during two distinct periods: Phase A (MJD 57040–57050) and Phase B (MJD 57065–57070). 
Despite the data we mentioned above, we collected simultaneously observed radio spectral data from \citet{Marina2018}, and the X-ray spectra were corrected through a de-absorption process employing the cross-section in \citet{Morrison1983ApJ} and the hydrogen column density value \citep{HI4PI2016}.
Historical archival data were also obtained from the Space Science Data Center (SSDC)\footnote{\url{https://www.ssdc.asi.it/}}.

Specifically, \citet{Wierzcholska2016MNRAS} performed cross-correlation analysis between the optical, UV, and $\gamma$-ray bands during January and February 2015. 
Their findings indicated no evident time lags among the optical, UV, and $\gamma$-ray emissions, suggesting they likely originate from the same region. 
The \textit{Swift}-XRT X-ray band is also expected to exhibit zero time lag with the $\gamma$-ray band; however, the mismatched sampling in the \textit{Swift}-XRT X-ray data could introduce artifacts affecting the correlation results \citep{Wierzcholska2016MNRAS}. 
Therefore, the one-zone model shall be considered during the SED modeling.
The popular method for estimating the (intrinsic) size of the emission region ($R_{\rm b}$) assumes that the flux variability timescale corresponds to the light travel time across the emission region. 
In this context, the observed shortest doubling/halving timescale ($t_{\rm var}$) can be used to constrain the size of the emission region, which is expressed as $R_{\rm b} \leqslant \frac{c t_{\rm var} \delta}{1+z}$ where $\delta$ is the Doppler factor. 
In our estimations, $t_{\rm var}$ was calculated as $t_{\rm var} = \ln(2) \times \min \left\{ T_{\rm r}, \, T_{\rm d} \right\}$ \citep[see, e.g.,][using the rising and decay components in Phase A for SED modeling]{Rani2013A&A557, Gasparyan2018ApJ863}. 
Following \citet{Bach2005A&A}, we adopted $\delta = 30$, leading to an estimated emission region size of $R_{\rm b} \simeq 4 \times 10^{16}$ cm during our SED modeling.
In the following, we performed the SED modeling using the public code \textit{Jets SED modeler and fitting Tool} \citep[\texttt{JetSet};][]{Tramacere2009A&A501, Tramacere2011ApJ739, Tramacere2020ascl09001}, and we considered two scenarios: (i) the broad-band emission originated from the leptonic model, namely synchrotron and IC radiation; and (ii) the emission originated from the lepto-hadronic hybrid model.

\subsubsection{Leptonic scenario}

The one-zone leptonic scenario assumes that the emissions originate from a spherical region (blob) of radius $R_{\rm b}$, filled with a uniform magnetic field ($B$). 
This region, located within the blazar jet, moves with a bulk Lorentz factor ($\Gamma = \frac{1}{\sqrt{1-\beta_{\Gamma}^{2}}} \sim \delta$, where $\beta_{\Gamma} c$ is the speed of the blob) at a small viewing angle to the observer, resulting in Doppler-boosted emission characterized by a Doppler factor ($\delta$).
Blazar emission is primarily dominated by radiation from synchrotron and IC processes. 
When the low-energy seed photons for the IC process originate from synchrotron radiation, the process is referred to as synchrotron self-Compton \citep[SSC; e.g.,][]{Finke2008ApJ686}. 
Alternatively, if the seed photons come from external regions such as the accretion disk \citep[AD;][]{Dermer1993ApJ}, the broad-line region \citep[BLR;][]{Sikora1994ApJ}, or the dust torus \citep[DT;][]{Blazejowski2000ApJ}, the process is termed external Compton (EC) process.

The blob is assumed to be filled with relativistic electrons with a log-parabola with low-energy power-law branch distribution, expressed as follows:
\begin{equation}
    N_{e}(\gamma) = A_{e} n_{e}(\gamma)= A_{e}
    \left\{ \begin{array}{l}
    \left( \frac{\gamma}{\gamma_{e, \, \rm 0}} \right) ^{-s},  ~~  \gamma_{e, \, \rm min} < \gamma < \gamma_{e, \, \rm 0} \\
    \left(\frac{\gamma}{\gamma_{e, \, \rm 0}} \right)^{- \left( s + r \log \left(\frac{\gamma}{\gamma_{e, \, \rm 0}}\right) \right)} ,   ~~   \gamma_{e, \, \rm 0} < \gamma < \gamma_{e, \, \rm max}
    \end{array}\right.  {\rm .}
\end{equation}
Here, $A_{e}$ is defined by the actual density of relativistic electrons ($N_{e}$) in units of $\rm cm^{-3}$ with $A_{e} = \frac{  N_{e} }{ \int n_{e}(\gamma) \mathrm{d}\gamma} $ (see the documentation in \texttt{JetSet}), 
$s$ is the spectral index, $r$ is the spectral curvature, and $\gamma_{e, \, \rm min/0/max}$ are the minimum, turn-over, and maximum electron Lorentz factor, respectively.

\begin{figure}
    \centering
    \includegraphics[width=6.5 in]{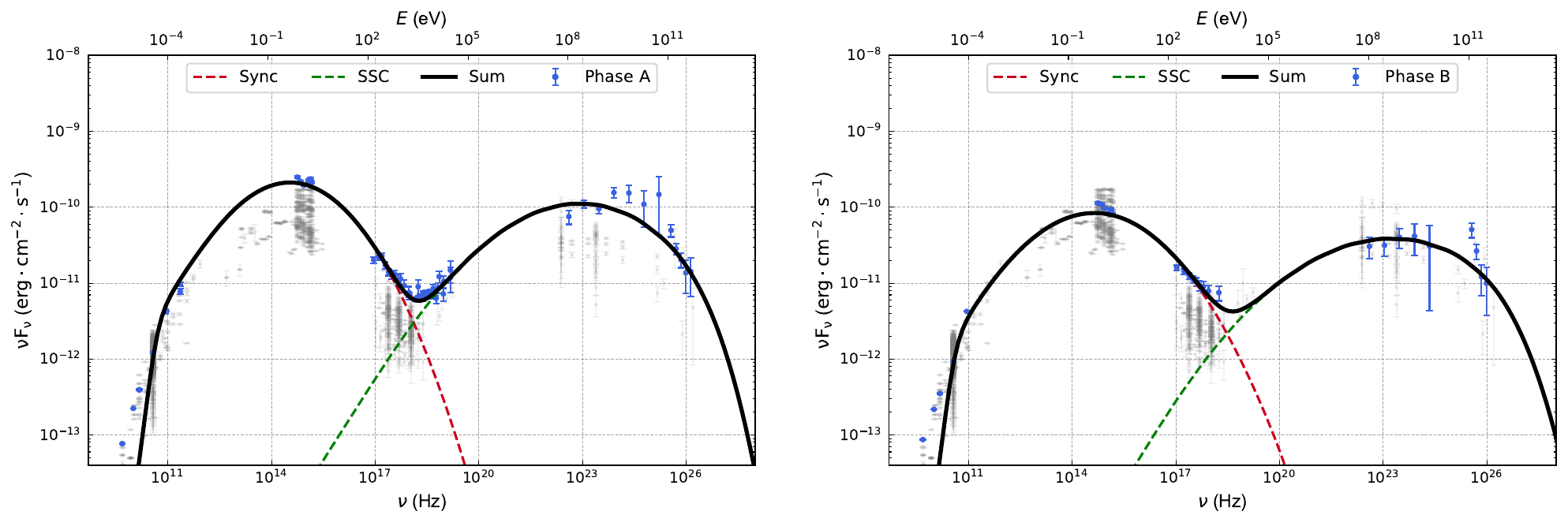}
    \caption{
    One-zone SSC modeling.
    The left panel is for Phase A (MJD 57040–57050); the right one is for Phase B (MJD 57065–57070). 
    The VHE spectra are corrected by EBL absorption adopting $z$=0.2304.  
    The meanings of line styles are given in the legend.
    }
    \label{fig:ssc_SED}
\end{figure}

S5 0716+714 is a BL Lac object, allowing us to model the SEDs using the one-zone SSC model.
The SSC model is described by nine parameters, where six of these parameters characterize the electron energy distribution ($\gamma_{e, \, \rm min/0/max}$, $N_{e}$, $s$, and $r$), while the remaining three describe the properties of the emission region ($R_{\rm b}$, $B$, and $\delta$). 
During the SSC modeling, the size of the emission region, $R_{\rm b}$, was fixed at 4 $\times$ 10$^{16}$ as mentioned above, and the minimum and maximum electron Lorentz factors were fixed at 1 and 10$^{7}$, respectively. 
The other parameters were optimized to achieve the best-fit model.
The best-fitted model parameters are summarized in Tab. \ref{tab:SED_params} and the resulting best-fit SSC model SEDs are presented in Fig. \ref{fig:ssc_SED}.

However, the SSC model failed to reproduce the large separation between the low-energy peak ($\sim 10^{14}$–$10^{15}$ Hz) and the high-energy peak ($\sim 10^{24}$–$10^{25}$ Hz) using a Doppler factor of $\delta = 30$ \citep{Bach2005A&A}. 
Such a large separation required an extreme Doppler factor ($\delta \sim 200$, see Tab. \ref{tab:SED_params}); otherwise, the observed optical flux would be underestimated. 
Moreover, a hard electron spectral index and a low magnetic field were also required. 
These difficulties could be avoided by assuming that the HE/VHE $\gamma$-ray emissions originate from a more highly boosted substructure within the jet.
For example, the ``jets-in-a-jet" model could have an extra Lorentz factor due to the plasma material outflowing from the reconnection regions \citep{Giannios2009MNRAS395L}. 
However, this scenario typically results in fast variability, as observed in sources like Mrk 501, PKS 2155-304, 3C 279, and M87 \citep{Albert2007ApJ669, Aharonian2007ApJ664L, Giannios2010MNRAS402, Shukla2020NatCo11}, which contradicts the variability observed in S5 0716+714.

One can calculate the jet power and further understand the composition of the jet. 
The jet power ($P_{\rm jet}$) carried by relativistic electrons ($P_{e}$), cold protons ($P_{p, \, \rm cold}$), and magnetic field ($P_{B}$) is estimated via
\begin{equation}
    P_{\rm jet} = \sum_{i} \pi R_{\rm b}^{2} \Gamma^{2} \beta_{\Gamma} c U_{i} {\rm ,}
    \label{P_jet}
\end{equation}
where the $U_{i}$ is the energy density of the relativistic electrons ($U_{e}$), cold protons ($U_{p, \, \rm cold}$), and magnetic field ($U_{B}$), respectively, in the co-moving frame \citep{Ghisellini2010MNRAS}. 
These energy densities can be derived by 
\begin{equation}
    U_{B} = \frac{B^{2}}{8 \pi}  {\rm ,}
\end{equation}
\begin{equation}
    U_{e} = m_{e} c^{2} \int \gamma N_{\rm e}(\gamma) \mathrm{d}\gamma  {\rm ,}
\end{equation}
\begin{equation}
    U_{p, \, \rm cold} = m_{p} c^{2} 1 \int   N_{e}(\gamma) \mathrm{d}\gamma  {\rm ,}
\end{equation}
where assuming a cold-proton-to-electron number density ratio of 1, $B$ is the magnetic field strength obtained from the SED modeling, $m_{e}$ and $m_{p}$ are the rest mass of the electron and proton, respectively. 
The jet powers and the energy densities are calculated and listed in Tab. \ref{tab:SED_params}. 
The Eddington luminosity of the supermassive black hole (SMBH) is 
\begin{equation}
    L_{\rm Edd} = 2 \pi m_{p} c^{3} R_{\rm S} / \sigma_{\rm T} {\rm ,}
\end{equation}
where $\sigma_{\rm T}$ is the Thompson scattering cross-section, $R_{\rm S} = 2 G M_{\rm BH} / c^{2}$ is the Schwarzschild radius of the SMBH, and $M_{\rm BH}$ is the mass of SMBH.
Using the $M_{\rm BH}$ = 10$^{8.91} M_{\odot}$ for S5 0716+714 \citep{Liu2019ApJ} where $M_{\odot}$ is the mass of the Sun, the Eddington luminosity of the source is calculated as $L_{\rm Edd} = 1.02 \times 10^{47}$ erg s$^{-1}$.  
It is worth noting that the jet powers in two phases under the SSC model moderately exceed the Eddington luminosity, primarily due to the extreme Doppler factors required to account for the SSC model.

As discussed above, the HE/VHE $\gamma$-ray emissions probably require a substructure within the jet to account for the large separation between the two peaks of the SEDs. 
Alternatively, these emissions could arise from an additional radiation component, such as the EC or hadronic component, which will be explored in the upcoming content and the next subsection, respectively.

\begin{figure}
    \centering
    \includegraphics[width=6.5 in]{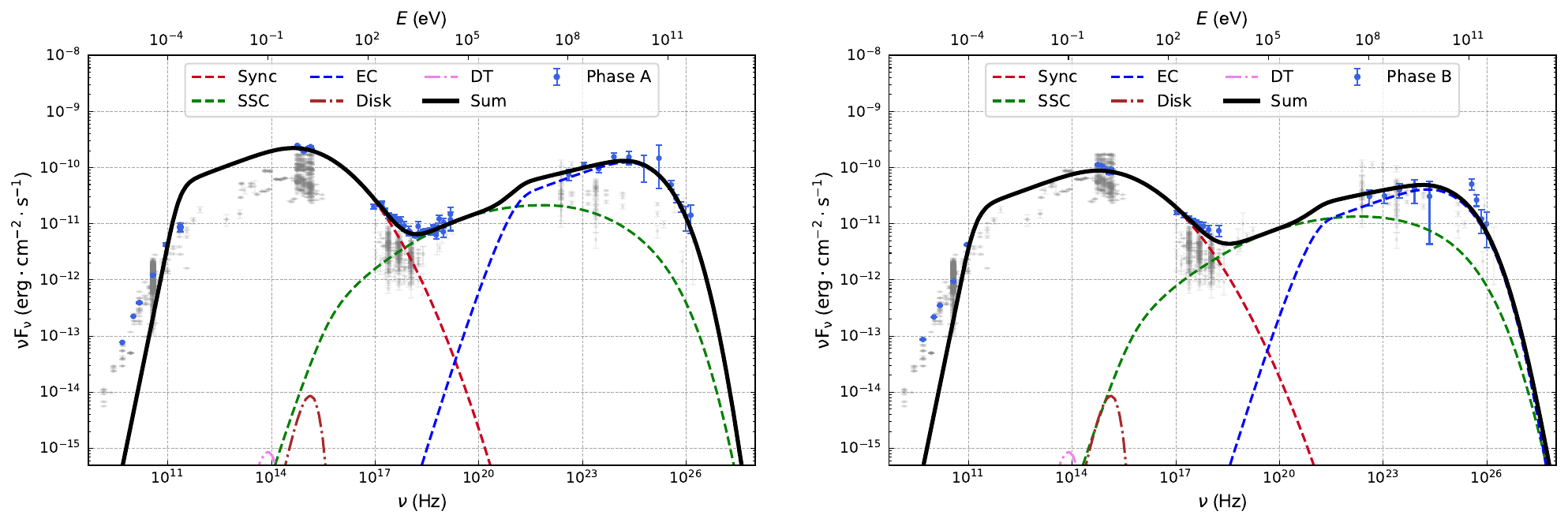}
    \caption{
    One-zone SSC plus EC modeling. 
    The left panel is for Phase A (MJD 57040–57050); the right one is for Phase B (MJD 57065–57070). 
    The VHE spectra are corrected by EBL absorption adopting $z$=0.2304. 
    The meanings of line styles are given in the legend, where the EC line includes the contribution from EC-BLR and EC-DT.
    }
    \label{fig:ssc+EC_SED}
\end{figure}

Previous studies have conducted SED modeling for S5 0716+714; however, they also found that the SSC model failed to reproduce the SEDs \citep[e.g.,][]{Tagliaferri2003A&A, Rani2013A&A552, Liao2014ApJ}. 
In \citet{Marina2018}, they attempted the one-zone SSC model but found it underestimated the observed optical flux.
Therefore, the EC component is a plausible mechanism for accounting for the HE/VHE $\gamma$-ray emissions, assuming the scattering of weak external emissions, despite the absence of detected thermal components or emission lines \citep{Shaw2009ApJ704, Liao2014ApJ}. 
Following this, we explored the SSC model combined with an EC component. 
Assuming a conical jet with a half-opening angle of $\theta_{\rm open}$ = 5$^{\circ}$, the distance of the emission region from the central engine ($R_{\rm H}$) was calculated to be $\sim$0.15 pc. 
We considered broad-line region as a spherical shell with an inner and outer radius of $R_{\rm BLR, \, in} = 3\times 10^{17} (L_{\rm disk}/10^{46})^{1/2}$ cm and 1.1$\times R_{\rm BLR, \, in}$ \citep{Kaspi2007ApJ659} with a coverage factor of $\tau_{\rm BLR}$ = 0.1, where $L_{\rm disk}$ is the accretion disk luminosity. 
The dust torus was assumed to be a radius of $R_{\rm DT} = 2\times 10^{19} (L_{\rm disk}/10^{46})^{1/2}$ cm \citep{Cleary2007ApJ660} with a reprocessing factor $\tau_{\rm DT}$ = 0.1. 
The dust torus and accretion disk temperatures were fixed at 1.2$\times$10$^{3}$ K and 2$\times$10$^{4}$ K, respectively. 
The actual disk luminosity ($L_{\rm disk}$) is challenging to constrain due to the featureless optical spectra of S5 0716+714. 
Therefore we fixed a reasonable disk luminosity value of $ 2 \times 10^{42}$ erg s$^{-1}$ during the SED modeling in two phases. 
This value was estimated according to the SED modeling and is below the upper limits reported in \citet{Ghisellini2010MNRAS} and \citet{Danforth2013ApJ}, making it a reasonable choice. 
Furthermore, the minimum electron Lorentz factor, $\gamma_{\rm e, \, min}$, was also left as a free parameter. 
The best-fit models are displayed in Fig. \ref{fig:ssc+EC_SED}, with the corresponding parameters listed in Tab. \ref{tab:SED_params}.

We can see that the SSC plus EC model provides a good fit to the SEDs and successfully reproduces the $\gamma$-ray emissions.
This is consistent with earlier studies, where the inclusion of EC components also successfully accounted for the SEDs \citep{Tagliaferri2003A&A, Rani2013A&A552, Liao2014ApJ}. 
The jet powers derived from this model using Eq. (\ref{P_jet}) are listed in Tab. \ref{tab:SED_params} and remain below the Eddington luminosity in both phases. 
This model requires a high Doppler factor of $\delta = 73.95$ for Phase A and $\delta = 57.84$ for Phase B, which are not preferred in such models for the causality of light traveling across the emission region.
While a higher Doppler factor corresponds to a larger variability timescale in the blob frame, such variability timescale could not be associated with the light-crossing but with the particle cooling processes or changes in external radiation fields.
Moreover, such high Doppler factors derived from the SED modeling remain problematic due to the maximum apparent velocity of $\beta_{\rm app}^{\rm max} \sim$ 34.4 estimated from \citet{Lister2018ApJS234}. 
Consequently, this model is excluded from our consideration.

\subsubsection{Lepto-hadronic hybrid scenario}

As mentioned above, the hadronic radiation component could be the origination to produce the $\gamma$-ray emissions.
The proton-proton (\textit{pp}) interactions have been used to explain the SEDs of blazars \citep[e.g.,][]{Banik2019PhRvD99, Banik2020PhRvD101}. 
In the work of \citet{Li2022A&A659}, they suggested that \textit{pp} interactions could be important for blazars and have a parameter space to interpret the $\gamma$-ray spectra.
Meanwhile, \citet{Xue2022PhRvD106} showed that \textit{pp} interactions can explain the TeV spectra. 
Therefore, we incorporated the \textit{pp} interactions into the one-zone SSC model to reproduce the SEDs following the parameters developed by \citet{Kelner2006PhRvD74}.
The \textit{pp} interactions will produce secondary neutral ($\pi^{0}$) and charged ($\pi^{\pm}$) pions, which will decay into electrons/positrons ($e^{\pm}$), neutrinos ($\nu$), and $\gamma$-ray emissions.
The \textit{pp} interactions are comprised of the following
\begin{equation}
 p + p \longrightarrow 
    \left\{ \begin{array}{l}
     p \\
     \pi^{0} \longrightarrow \gamma + \gamma \\
     \pi^{+} \longrightarrow \mu^{+} + \nu_{\mu} \longrightarrow e^{+} + \nu_{e} + \bar \nu_{\mu} + \nu_{\mu} \\ 
     \pi^{-} \longrightarrow \mu^{-} + \bar \nu_{\mu} \longrightarrow e^{-} + \bar \nu_{e} + \nu_{\mu} + \bar \nu_{\mu}
    \end{array}\right. { \rm .}
\end{equation}

We adopted the analytical results from \citet{Li2022A&A659}, which indicate that if the $\gamma$-ray emissions originate from $\pi^{0}$ decay and the jet power does not exceed the Eddington luminosity, the size of the emission region is constrained by
\begin{equation}
    \frac{R_{\rm b}}{R_{\rm S}} \leqslant \frac{\sigma_{pp}}{12 \sigma_{\rm T}} \frac{L_{\rm Edd}}{L_{\rm TeV}^{\rm obs}}  {\rm ,}
    \label{pp_R}
\end{equation}
where $\sigma_{pp} \approx$ 6$\times$10$^{-26}$ cm$^{2}$ is the cross-section for \textit{pp} interactions \citep{Kelner2006PhRvD74}, and $L_{\rm TeV}^{\rm ob}$ is the observed TeV luminosity.
Applying the method from \citet{Xue2019ApJ871} to calculate $L_{\rm TeV}^{\rm ob}$, we found that the maximum allowable $R_{\rm b}$ is constrained to $\sim 10^{13}$ cm.
This implies a very compact emission region, smaller than the Schwarzschild radius ($R_{\rm S}$ = 2.44$\times$10$^{14}$ cm) of S5 0716+714, which is atypical for blazars.
Consequently, the jet power somewhat exceeding the Eddington luminosity appears necessary, which is plausible during a flaring state \citep{Gao2019NatAs3, Banik2019PhRvD99, Banik2020PhRvD101}. 
At the same time, Equation (11) in \citet{Li2022A&A659}, which estimates the number density of cold protons ($n_{\rm H}$), was deemed inapplicable here.

\begin{figure}
    \centering
    \includegraphics[width=6.5 in]{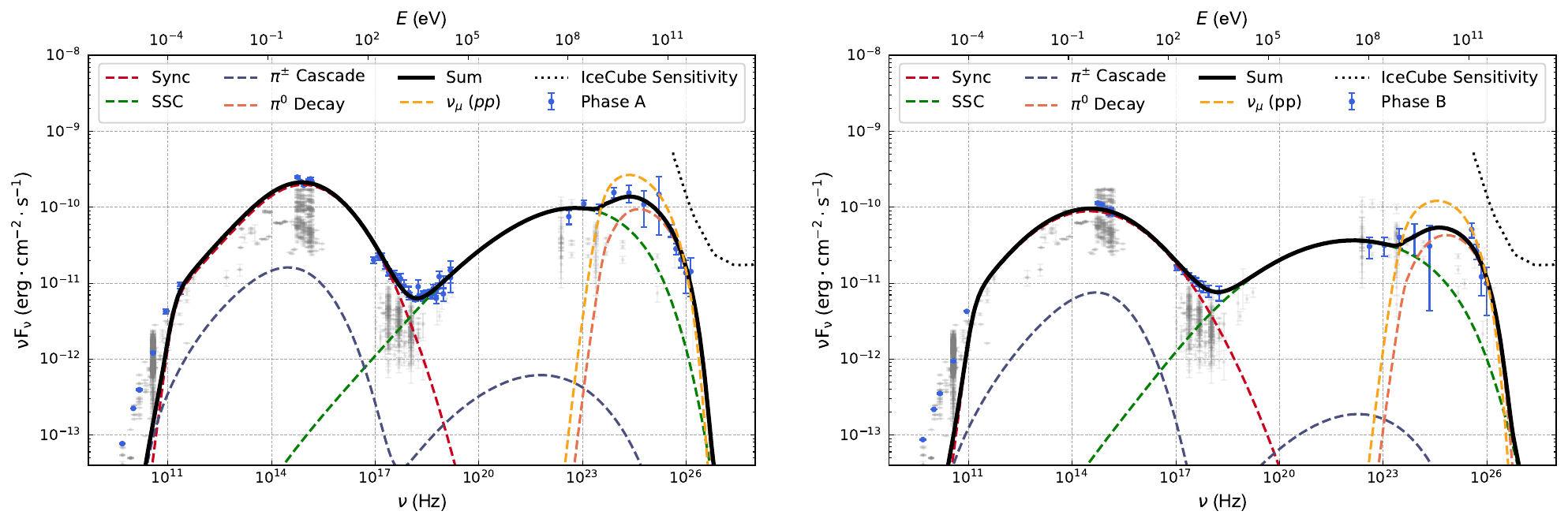}
    \caption{
    One-zone SSC plus \textit{pp} modeling. 
    The left panel is for Phase A (MJD 57040–57050); the right one is for Phase B (MJD 57065–57070). 
    The VHE spectra are corrected by EBL absorption adopting $z$=0.2304. 
    The meanings of line styles are given in the legend. 
    The dashed yellow line is the expected muon neutrino flux produced by the $\pi^{\pm}$ cascade through \textit{pp} interactions. 
    The dotted black line represents the IceCube sensitivity for declination $\delta_{\rm del}$=60 using the 10 yrs dataset from \citet{Ghiassi1781237}.
    }
    \label{fig:ssc+pp_SED}
\end{figure}

We assumed that the relativistic protons follow a power-law with an exponential cut-off distribution:
\begin{equation}
    N_{p} (\gamma) = A_{p} n_{p}(\gamma) = A_{p} \gamma^{- \alpha_{p}} \exp \left( -\frac{\gamma}{\gamma_{p, \, \rm cut}} \right), ~ \gamma_{p, \, \rm min} < \gamma < \gamma_{p, \, \rm max}  {\rm .}
\end{equation}
Here, $A_{p} = N_{p} \frac{ 1 }{ \int n_{p}(\gamma) \mathrm{d}\gamma} $, where $N_{p}$ refers to the actual density of relativistic protons in units of $\rm cm^{-3}$.
The parameter $\alpha_{p}$ is the power-law spectral index, $\gamma_{p, \, \rm min/cut/max}$ are the minimum, exponential cut-off, and maximum proton Lorentz factors, respectively.
The efficiency of \textit{pp} interactions ($f_{pp}$) depends on the density of cold protons and can be estimated using the expression: 
\begin{equation}
    f_{pp} = K_{pp} \sigma_{pp} n_{\rm H} R_{\rm b} {\rm ,}
\end{equation}
where $K_{pp} \approx$ 0.5 is the inelasticity coefficient, $n_{\rm H}$ is the number density of cold protons \citep{Kelner2006PhRvD74}. 
We set the number density of cold protons $n_{\rm H}$ = 10$^{4}$ cm$^{-3}$ and the proton Lorentz factors $\gamma_{p, \, \rm min/cut/max}$ = 1/10$^{2}$/10$^{3}$ in our modeling.
The cold proton column density would be $N_{\rm H, cold} \simeq n_{\rm H} R_{\rm b} \simeq$ 4$\times$10$^{20}$ cm$^{-2}$, making the optical/UV and X-ray emission absorbed via the photoionization absorption process.
The optical depth for scattering is $\tau_{\rm sc} = \sigma_{\rm sc} n_{\rm H} R_{\rm b}$ where $\sigma_{\rm sc}$ is the scattering cross-section, expressed by
\begin{equation}
    \sigma_{\rm sc} = \sigma_{\rm T} \frac{3}{4} \left[ \frac{1+x}{x^3} \left\{ \frac{2x(1+x)}{1+2x} - \ln(1+2x) \right\} + \frac{1}{2x} \ln(1+2x) - \frac{1+3x}{(1+2x)^2} \right] {\rm ,}
\end{equation}
with $x = E/(m_{e}c^{2})$ and $E$ represents the photon energy \citep[see][]{Liu2019PhRvD99}. 
The resulting flux should be multiplied by a factor of $\left( 1 - \exp(-\tau_{\rm sc}) \right) / \tau_{\rm sc}$. 
However, the absorption effect is negligible since $\tau_{\rm sc} \ll 1$ was obtained.
During the modeling, the Doppler factor was fixed at 30, the minimum electron Lorentz factor set to 1, while the remaining parameters were treated as free. 
The fitting results are presented in Fig. \ref{fig:ssc+pp_SED} with the corresponding parameters listed in Tab. \ref{tab:SED_params}.

The results indicate that the $\gamma$-ray emissions generated by $\pi^{0}$ decay via \textit{pp} interactions successfully reproduce the SEDs.
Additionally, the synchrotron emission from secondary electron pairs contributes marginally to the radio and X-ray fluxes compared to the emission from primary electrons.
The energy densities of the cold protons and relativistic protons in the SSC plus \textit{pp} model are calculated as follows:
\begin{equation}
U_{p, \, \rm cold} = m_{p} c^{2} n_{\rm H} {\rm ,}
\end{equation}
\begin{equation}
U_{p, \, \rm rel} = m_{p} c^{2} \int \gamma N_{p}(\gamma) \mathrm{d}\gamma {\rm .}
\end{equation}
The total jet power is estimated to be $P_{\rm jet}$ = 4.11 $\times$ 10$^{48}$ erg s$^{-1}$ for Phase A and $P_{\rm jet}$ = 3.09 $\times$ 10$^{48}$ erg s$^{-1}$ for Phase B, which exceed the Eddington luminosity by a factor of 30-40, as detailed in Tab. \ref{tab:SED_params}. 
An enhanced activity may result in a transient increase in jet power, potentially exceeding the Eddington luminosity, as discussed above. 
Alternatively, in the case of a highly collimated jet outflow, the Eddington luminosity can be exceeded because the jet does not interfere with the accretion flow \citep{Gao2019NatAs3}.
Moreover, VLBI analysis by \citet{Marina2018} suggests that a superluminal knot passing through a recollimation shock may occur during this flaring state.
Such interactions may increase the number density of electrons and protons in the emission region, further driving the jet power to exceed the Eddington luminosity.

The dashed yellow line in Fig. \ref{fig:ssc+pp_SED} shows the muon neutrino ($\nu_{\mu}$) flux produced by $\pi^{\pm}$ cascade. 
It is necessary to evaluate the possible neutrino emission under the hadronic model.
The corresponding neutrino event rate can be estimated via
\begin{equation}
    \frac{\mathrm{d} N_{\nu_{\mu}}}{\mathrm{d} t} = \int _{E_{\nu_{\mu}, \, \rm min}} ^{E_{\nu_{\mu}, \, \rm max}} \mathrm{d} E_{\nu_{\mu}} A_{\rm eff}(E_{\nu_{\mu}}, \delta_{\rm del}) \phi_{E_{\nu_{\mu}}}  
    {\rm ,}
\end{equation}
where ${E_{\nu_{\mu}, \, \rm min}}$ and ${E_{\nu_{\mu}, \, \rm max}}$ are the lower and upper limits of the
neutrino energy, respectively, $A_{\rm eff}(E_{\nu_{\mu}}, \delta_{\rm del})$ is the effective area in given declination ($\delta_{\rm del}$), and $\phi_{E_{\nu_{\mu}}}$ is the muon neutrino differential energy flux. 
Then the expected neutrino event rates for S5 0716+714, using the effective area from \citet{Carver2019ICRC36_851}, are 0.69 events yr$^{-1}$ for Phase A and 0.52 events yr$^{-1}$ for Phase B, respectively. 
However, the IceCube did not detect any neutrino events from S5 0716+715, because the expected neutrino flux was below its cumulative sensitivity, making it insufficient to detect such events from this source, as shown in the dotted black line in Fig. \ref{fig:ssc+pp_SED}.

In addition to the models mentioned above, the two-zone SSC model was also explored in \citet{Marina2018}, which considers the interaction between a superluminal knot and a recollimation shock within the jet. 
While this two-zone model provides a better fit to the SED, it cannot fully reproduce the observations in the 10-100 GeV range. 
Furthermore, other jet models, such as the structured jet model \citep{Marina2018} and the helical magnetic field model \citep{Chandra2015ApJ}, have also been investigated during this flaring period.
These various models indicate the complexity of the emission mechanisms during flaring states, suggesting that further multi-wavelength observations are needed to fully capture the underlying processes.

\begin{deluxetable}{cccccccc}
\tablewidth{12pt} 
\tablecaption{Parameters for the broadband SEDs for different models
\label{tab:SED_params}
}
\tablehead{ 
    \multirow{2}[2]{*}{Params} & \multirow{2}[2]{*}{Units} & \multicolumn{3}{c}{Phase A} & \multicolumn{3}{c}{Phase B} \\ \cmidrule(lr){3-5} \cmidrule(lr){6-8}
        &     & SSC   & SSC+EC & SSC+\textit{pp} interactions & SSC   & SSC+EC & SSC+\textit{pp} interactions
}
\colnumbers 
\startdata
    $R_{\rm b}$ $^{\dagger}$ & cm    & 4.00$\times$10$^{16}$ & 4.00$\times$10$^{16}$ & 4.00$\times$10$^{16}$ & 4.00$\times$10$^{16}$ & 4.00$\times$10$^{16}$ & 4.00$\times$10$^{16}$ \\
    $B$   & G     & 4.45$\times$10$^{-3}$ & 8.35$\times$10$^{-2}$ & 1.95$\times$10$^{-1}$ & 3.78$\times$10$^{-3}$ & 7.10$\times$10$^{-2}$ & 1.61$\times$10$^{-1}$ \\
    $\delta$ &       & 204.44 & 73.95 & 30 $^{\dagger}$    & 188.40 & 57.84 & 30 $^{\dagger}$ \\
    $\gamma_{e, \, \rm min}$ &       & 1 $^{\dagger}$     & 77.44 & 1 $^{\dagger}$     & 1 $^{\dagger}$     & 96.69 & 1 $^{\dagger}$ \\
    $\gamma_{e, \, \rm max}$ $^{\dagger}$ &       & 10$^{7}$ & 10$^{7}$ & 10$^{7}$ & 10$^{7}$ & 10$^{7}$ & 10$^{7}$ \\
    $N_{e}$ & $\rm cm^{-3}$ & 123.81 & 25.37 & 2071.33 & 104.12 & 23.85 & 1674.66 \\
    $\gamma_{e, \, 0}$ &       & 911.98 & 2540.8 & 1863.48 & 414.73 & 2961.56 & 329.75 \\
    $s$   &       & 1.57  & 2.54  & 1.99  & 1.52  & 2.55  & 1.85 \\
    $r$   &       & 0.64  & 0.78  & 0.87  & 0.47  & 0.59  & 0.48 \\
 \cmidrule(lr){1-8}
    $\tau_{\rm BLR}$ $^{\dagger}$ &       & --    & 0.1   & --    & --    & 0.1   & -- \\
    $T_{\rm DT}$ $^{\dagger}$ & K     & --    & 1200  & --    & --    & 1200  & -- \\
    $\tau_{\rm DT}$ $^{\dagger}$ &       & --    & 0.1   & --    & --    & 0.1   & -- \\
    $L_{\rm Disk}$ $^{\dagger}$ & $\rm erg \, s^{-1}$ & --    & 2$\times$10$^{42}$ & --    & --    & 2$\times$10$^{42}$ & -- \\
    $T_{\rm Disk}$ $^{\dagger}$ & K     & --    & 2$\times$10$^{4}$ & --    & --    & 2$\times$10$^{4}$ & -- \\
    $\theta_{\rm open}$ $^{\dagger}$ & deg   & --    & 5     & --    & --    & 5     & -- \\
    $R_{\rm H}$ $^{\dagger}$ & cm    & --    & 4.57$\times$10$^{17}$ & --    & --    & 4.57$\times$10$^{17}$ &  \\
 \cmidrule(lr){1-8}
    $\gamma_{p, \, \rm min}$ $^{\dagger}$ &       & --    & --    & 1     & --    & --    & 1 \\
    $\gamma_{p, \, \rm max}$ $^{\dagger}$ &       & --    & --    & 1000  & --    & --    & 1000 \\
    $N_{p}$ & $\rm cm^{-3}$ & --    & --    & 3185.17 & --    & --    & 1318.35 \\
    $\gamma_{p, \, \rm cut}$ $^{\dagger}$ &       & --    & --    & 100   & --    & --    & 100 \\
    $\alpha_{p}$ &       & --    & --    & 2.25  & --    & --    & 2.07 \\
    $n_{\rm H}$ $^{\dagger}$ & $\rm cm^{-3}$ & --    & --    & 10$^{4}$ & --    & --    & 10$^{4}$ \\
    $f_{pp}$ &   & --    & --    & 1.20$\times$10$^{-5}$ & --    & --    & 1.20$\times$10$^{-5}$ \\
 \cmidrule(lr){1-8}
    $U_{e}$ & $\rm erg \, cm^{-3}$ & 5.60$\times$10$^{-3}$ & 4.26$\times$10$^{-3}$ & 1.60$\times$10$^{-2}$ & 4.79$\times$10$^{-3}$ & 4.97$\times$10$^{-3}$ & 1.72$\times$10$^{-2}$ \\
    $U_{p, \, \rm cold}$ & $\rm erg \, cm^{-3}$ & 1.86$\times$10$^{-1}$ & 3.81$\times$10$^{-2}$ & 1.50$\times$10$^{1}$ & 1.57$\times$10$^{-1}$ & 3.59$\times$10$^{-2}$ & 1.50$\times$10$^{1}$ \\
    $U_{B}$ & $\rm erg \, cm^{-3}$ & 7.89$\times$10$^{-7}$ & 2.77$\times$10$^{-4}$ & 1.51$\times$10$^{-3}$ & 5.68$\times$10$^{-7}$ & 2.01$\times$10$^{-4}$ & 1.03$\times$10$^{-3}$ \\
 \cmidrule(lr){1-8}
    $P_{e}$ & $\rm erg \, s^{-1}$ & 3.53$\times$10$^{46}$ & 3.51$\times$10$^{45}$ & 2.16$\times$10$^{45}$ & 2.56$\times$10$^{46}$ & 2.51$\times$10$^{45}$ & 2.33$\times$10$^{45}$ \\
    $P_{B}$ & $\rm erg \, s^{-1}$ & 4.97$\times$10$^{42}$ & 2.28$\times$10$^{44}$ & 2.05$\times$10$^{44}$ & 3.04$\times$10$^{42}$ & 1.01$\times$10$^{44}$ & 1.39$\times$10$^{44}$ \\
    $P_{p, \, \rm cold}$ & $\rm erg \, s^{-1}$ & 1.17$\times$10$^{48}$ & 3.14$\times$10$^{46}$ & 2.04$\times$10$^{48}$ & 8.37$\times$10$^{47}$ & 1.81$\times$10$^{46}$ & 2.04$\times$10$^{48}$ \\
 \cmidrule(lr){1-8}
    $U_{\rm BLR}$ & $\rm erg \, cm^{-3}$ & --    & 2.70$\times$10$^{-10}$ & --    & --    & 2.93$\times$10$^{-10}$ & -- \\
    $U_{\rm DT}$ & $\rm erg \, cm^{-3}$ & --    & 1.84$\times$10$^{-4}$ & --    & --    & 1.13$\times$10$^{-4}$ & -- \\
 \cmidrule(lr){1-8}
    $U_{p, \, \rm rel}$ & $\rm erg \, cm^{-3}$ & --    & --    & 1.52$\times$10$^{1}$ & --    & --    & 7.70$\times$10$^{0}$ \\
    $P_{p, \, \rm rel}$ & $\rm erg \, s^{-1}$ & --    & --    & 2.07$\times$10$^{48}$ & --    & --    & 1.04$\times$10$^{48}$ \\
 \cmidrule(lr){1-8}
    $P_{\rm tot}$ & $\rm erg \, s^{-1}$ & 1.21$\times$10$^{48}$ & 3.52$\times$10$^{46}$ & 4.11$\times$10$^{48}$ & 8.63$\times$10$^{47}$ & 2.07$\times$10$^{46}$ & 3.09$\times$10$^{48}$ \\
\enddata
\tablecomments{ 
Parameters with the symbol ``${\dagger}$" represent that they keep frozen in the SED fitting. 
The redshift is adopted as 0.2304. 
The symbol ``--" represents a null value. 
The Eddington luminosity is calculated as $L_{\rm Edd} = 1.02\times10^{47}$ erg s$^{-1}$ for the SMBH mass 10$^{8.91} M_{\odot}$ for S5 0716+714 \citep{Liu2019ApJ}.  
}
\end{deluxetable}

\section{Summary} \label{Summary}

In this study, we conducted a multi-wavelength analysis of the 2015 flare of S5 0716+714 to investigate its radiation mechanisms. 
The data were gathered from the \textit{Swift}-UVOT, \textit{Swift}-XRT, \textit{NuSTAR}, and \textit{Fermi}-LAT databases, along with the MAGIC VHE data from the literature. 
These observations allowed us to estimate the size of the emission region, while the modeling of the broad-band SEDs for Phase A and Phase B provided valuable insights into the physical processes occurring during the flaring periods.
The main results are as follows. 
The size of the $\gamma$-ray emission region was estimated using the variability timescale, determined through exponential function fitting. 
Subsequently, one-zone models, including leptonic and lepto-hadronic hybrid scenarios, were employed to reproduce the SEDs for Phase A and Phase B. 
However, the SSC models could not adequately describe the SEDs without invoking extreme Doppler factors, potentially requiring an additional Lorentz factor as suggested in the ``jets-in-a-jet" model.
The SSC plus EC model provided a good fit to the SEDs but required a high Doppler factor, leading to its exclusion from our consideration.
Additionally, the SSC plus \textit{pp} interactions model was explored, and the results demonstrated that this model successfully reproduced the SEDs. 
Nevertheless, the total jet power in this scenario exceeded the Eddington luminosity, a situation that is still plausible due to the flaring state or the presence of a highly collimated jet.

\begin{acknowledgments}


H.B.X. acknowledges the support from the National Natural Science Foundation of China (NSFC 12203034), the Shanghai Science and Technology Fund (22YF1431500), and the science research grants from the China Manned Space Project.
MM acknowledges the Croatian Science Foundation (HrZZ) Project IP-2022-10-4595.
R.X acknowledges the support from the NSFC under grant No. 12203043.
G.W. acknowledges the support from the China Postdoctoral Science Foundation (grant No. 2023M730523).
S.H.Z acknowledges support from the National Natural Science Foundation of China (Grant No. 12173026), the National Key Research and Development Program of China (Grant No. 2022YFC2807303), the Shanghai Science and Technology Fund (Grant No. 23010503900), the Program for Professor of Special Appointment (Eastern Scholar) at Shanghai Institutions of Higher Learning and the Shuguang Program (23SG39) of the Shanghai Education Development Foundation and Shanghai Municipal Education Commission.
J.H.F acknowledges the support from the NSFC 12433004, NSFC U2031201,  the Scientific and Technological Cooperation Projects (2020–2023) between the People’s Republic of China and the Republic of Bulgaria, the science research grants from the China Manned Space Project with No. CMS-CSST-2021-A06, and the support for Astrophysics Key Subjects of Guangdong Province and Guangzhou City.
This research was partially supported by the Bulgarian National Science Fund of the Ministry of Education and Science under grants KP-06-H38/4 (2019), KP-06-KITAJ/2 (2020) and KP-06-H68/4 (2022).

\end{acknowledgments}



\appendix

\section{\textit{Swift} observation data}

The \textit{Swift}-XRT best-fit parameters with an absorbed power-law model are shown in Tab. \ref{tab:XRT}. 

\begin{deluxetable}{cccccccccccc}
\tabletypesize{\scriptsize}
\tablewidth{0pt} 
\tablecaption{\textit{Swift}-XRT observations \label{tab:XRT}}
\tablehead{ 
    ObsID & Mode  & Exposure & Time & $\Gamma_{\rm X}$ & $\Gamma_{\rm X, \, err}$ & $N_{0}$ & $N_{0, \, \rm err}$ & Flux & Flux$_{\rm err}$ & C-statistic & d.o.f. 
}
\colnumbers 
\startdata
    00035009143 & PC    & 921.50 & 57011.04 & 2.18  & 0.23  & 1.08E-03 & 1.62E-04 & 5.60E-12 & 9.57E-13 & 72.59 & 91 \\
    00035009144 & PC    & 1066.34 & 57019.35 & 1.78  & 0.18  & 1.32E-03 & 1.71E-04 & 8.61E-12 & 1.63E-12 & 142.18 & 137 \\
    00035009145 & PC    & 1051.36 & 57023.21 & 2.26  & 0.18  & 1.55E-03 & 1.82E-04 & 7.80E-12 & 1.00E-12 & 99.42 & 138 \\
    00035009146 & PC    & 829.10 & 57029.01 & 1.81  & 0.22  & 1.27E-03 & 1.88E-04 & 8.08E-12 & 1.78E-12 & 119.58 & 105 \\
    00035009147 & PC    & 991.42 & 57041.09 & 2.22  & 0.12  & 4.08E-03 & 3.01E-04 & 2.08E-11 & 1.72E-12 & 187.33 & 217 \\
    00035009148 & PC    & 1103.80 & 57042.75 & 2.11  & 0.11  & 3.93E-03 & 2.92E-04 & 2.10E-11 & 1.87E-12 & 199.73 & 230 \\
    00035009149 & PC    & 968.95 & 57043.41 & 2.05  & 0.15  & 4.01E-03 & 3.91E-04 & 2.19E-11 & 2.65E-12 & 156.73 & 179 \\
    00035009152 & PC    & 1356.03 & 57044.02 & 2.29  & 0.09  & 5.85E-03 & 3.53E-04 & 2.93E-11 & 1.91E-12 & 224.78 & 259 \\
    00035009153 & PC    & 6912.53 & 57044.29 & 2.19  & 0.06  & 4.98E-03 & 1.90E-04 & 2.57E-11 & 1.12E-12 & 349.31 & 382 \\
    00035009154 & PC    & 998.94 & 57045.01 & 2.31  & 0.11  & 7.10E-03 & 4.85E-04 & 3.53E-11 & 2.57E-12 & 168.92 & 228 \\
    00035009156 & PC    & 9574.62 & 57045.14 & 2.43  & 0.04  & 7.95E-03 & 2.19E-04 & 3.87E-11 & 1.09E-12 & 425.84 & 425 \\
    00035009157 & PC    & 1688.17 & 57047.14 & 2.42  & 0.09  & 9.34E-03 & 5.32E-04 & 4.55E-11 & 2.65E-12 & 211.43 & 249 \\
    00035009158 & PC    & 6557.90 & 57047.22 & 2.46  & 0.05  & 9.28E-03 & 2.67E-04 & 4.50E-11 & 1.31E-12 & 352.41 & 391 \\
    00035009159 & PC    & 1490.88 & 57048.73 & 2.38  & 0.10  & 8.14E-03 & 5.15E-04 & 3.99E-11 & 2.62E-12 & 211.88 & 226 \\
    00035009160 & PC    & 1490.88 & 57048.86 & 2.41  & 0.10  & 6.32E-03 & 3.86E-04 & 3.08E-11 & 1.93E-12 & 189.58 & 238 \\
    00035009161 & PC    & 1490.88 & 57049.66 & 2.28  & 0.08  & 6.26E-03 & 3.31E-04 & 3.14E-11 & 1.79E-12 & 247.94 & 285 \\
    00035009162 & WT    & 1007.11 & 57050.01 & 2.73  & 0.12  & 5.77E-03 & 4.11E-04 & 2.81E-11 & 2.05E-12 & 222.56 & 247 \\
    00035009167 & PC    & 3161.58 & 57051.26 & 2.02  & 0.06  & 3.73E-03 & 1.65E-04 & 2.07E-11 & 1.15E-12 & 328.36 & 366 \\
    00035009164 & WT    & 198.21 & 57051.66 & 2.37  & 0.37  & 4.12E-03 & 8.68E-04 & 2.02E-11 & 4.47E-12 & 60.00 & 87 \\
    00035009168 & PC    & 2469.83 & 57051.85 & 2.09  & 0.07  & 3.50E-03 & 1.76E-04 & 1.88E-11 & 1.15E-12 & 291.47 & 329 \\
    00035009169 & WT    & 5488.26 & 57052.27 & 2.25  & 0.06  & 4.82E-03 & 1.58E-04 & 2.43E-11 & 8.90E-13 & 472.61 & 515 \\
    00035009170 & WT    & 6144.92 & 57052.99 & 2.21  & 0.06  & 4.36E-03 & 1.39E-04 & 2.23E-11 & 8.23E-13 & 548.34 & 553 \\
    00035009171 & WT    & 5502.71 & 57054.46 & 2.29  & 0.05  & 5.15E-03 & 1.47E-04 & 2.57E-11 & 8.00E-13 & 471.24 & 518 \\
    00035009172 & WT    & 1082.94 & 57055.45 & 2.37  & 0.11  & 5.15E-03 & 3.38E-04 & 2.53E-11 & 1.74E-12 & 242.50 & 266 \\
    00035009173 & WT    & 1072.83 & 57056.31 & 2.51  & 0.07  & 1.08E-02 & 4.64E-04 & 5.20E-11 & 2.25E-12 & 301.62 & 329 \\
    00035009174 & PC    & 1483.39 & 57057.64 & 2.34  & 0.08  & 7.01E-03 & 3.70E-04 & 3.47E-11 & 1.93E-12 & 262.78 & 269 \\
    00035009175 & WT    & 2069.23 & 57058.52 & 2.45  & 0.06  & 7.57E-03 & 2.84E-04 & 3.68E-11 & 1.40E-12 & 348.40 & 384 \\
    00035009176 & WT    & 5406.15 & 57058.65 & 2.46  & 0.04  & 8.62E-03 & 1.90E-04 & 4.18E-11 & 9.35E-13 & 494.67 & 549 \\
    00035009177 & WT    & 13689.18 & 57059.05 & 2.49  & 0.02  & 8.46E-03 & 1.17E-04 & 4.10E-11 & 5.69E-13 & 739.93 & 696 \\
    00035009178 & WT    & 12428.53 & 57060.05 & 2.47  & 0.03  & 7.29E-03 & 1.21E-04 & 3.53E-11 & 5.93E-13 & 636.68 & 682 \\
    00035009179 & WT    & 12276.77 & 57061.04 & 2.41  & 0.03  & 6.65E-03 & 1.17E-04 & 3.25E-11 & 5.86E-13 & 602.31 & 683 \\
    00035009180 & WT    & 16028.34 & 57062.05 & 2.42  & 0.03  & 5.85E-03 & 9.51E-05 & 2.85E-11 & 4.75E-13 & 605.02 & 701 \\
    00035009181 & WT    & 1122.72 & 57063.96 & 2.39  & 0.11  & 5.01E-03 & 3.52E-04 & 2.45E-11 & 1.78E-12 & 206.55 & 252 \\
    00035009182 & WT    & 978.13 & 57064.30 & 2.60  & 0.13  & 5.06E-03 & 3.92E-04 & 2.44E-11 & 1.89E-12 & 197.69 & 228 \\
    00035009184 & WT    & 648.52 & 57066.03 & 2.27  & 0.14  & 6.74E-03 & 5.79E-04 & 3.39E-11 & 3.18E-12 & 202.84 & 223 \\
    00035009185 & WT    & 4604.37 & 57066.09 & 2.46  & 0.04  & 7.65E-03 & 1.89E-04 & 3.72E-11 & 9.29E-13 & 484.02 & 501 \\
    00035009186 & WT    & 4470.44 & 57066.10 & 2.47  & 0.05  & 7.80E-03 & 2.06E-04 & 3.78E-11 & 1.01E-12 & 463.70 & 486 \\
    00035009187 & WT    & 998.11 & 57068.29 & 2.55  & 0.10  & 6.77E-03 & 3.98E-04 & 3.27E-11 & 1.92E-12 & 225.36 & 273 \\
    00035009188 & WT    & 831.23 & 57068.82 & 2.66  & 0.15  & 4.39E-03 & 3.79E-04 & 2.12E-11 & 1.85E-12 & 186.74 & 212 \\
    00035009189 & WT    & 1028.06 & 57069.95 & 2.40  & 0.14  & 3.91E-03 & 3.16E-04 & 1.91E-11 & 1.60E-12 & 209.45 & 228 \\
    00035009190 & PC    & 1475.90 & 57067.91 & 2.34  & 0.10  & 7.71E-03 & 4.81E-04 & 3.81E-11 & 2.50E-12 & 218.48 & 243 \\
    00035009192 & WT    & 1077.83 & 57070.23 & 2.46  & 0.18  & 3.75E-03 & 3.81E-04 & 1.82E-11 & 1.88E-12 & 200.21 & 207 \\
    00035009193 & WT    & 1097.78 & 57070.75 & 2.37  & 0.14  & 3.48E-03 & 2.92E-04 & 1.71E-11 & 1.49E-12 & 202.51 & 228 \\
    00035009194 & WT    & 878.14 & 57071.09 & 2.47  & 0.14  & 4.06E-03 & 3.47E-04 & 1.97E-11 & 1.70E-12 & 181.57 & 222 \\
    00035009195 & WT    & 1062.83 & 57071.69 & 2.48  & 0.13  & 4.25E-03 & 3.23E-04 & 2.06E-11 & 1.58E-12 & 232.91 & 242 \\
    00035009196 & PC    & 1018.89 & 57072.01 & 2.28  & 0.13  & 4.08E-03 & 3.37E-04 & 2.05E-11 & 1.84E-12 & 147.91 & 189 \\
    00035009197 & PC    & 476.99 & 57072.75 & 2.21  & 0.17  & 3.95E-03 & 4.34E-04 & 2.02E-11 & 2.50E-12 & 108.49 & 145 \\
    00035009198 & PC    & 1016.40 & 57073.01 & 2.31  & 0.12  & 4.28E-03 & 3.40E-04 & 2.13E-11 & 1.81E-12 & 136.03 & 198 \\
    00035009199 & PC    & 1111.29 & 57073.61 & 2.18  & 0.10  & 4.15E-03 & 2.85E-04 & 2.14E-11 & 1.68E-12 & 223.43 & 237 \\
    00035009200 & PC    & 1016.40 & 57074.14 & 2.19  & 0.12  & 3.53E-03 & 2.75E-04 & 1.82E-11 & 1.62E-12 & 210.63 & 208 \\
    00035009201 & PC    & 1073.83 & 57074.61 & 2.17  & 0.14  & 2.83E-03 & 2.59E-04 & 1.47E-11 & 1.55E-12 & 149.23 & 175 \\
\enddata
\tablecomments{
Col. (1): the ObsID; 
Col. (2): the readout mode, PC represents the Photon Counting mode and WT represents the Windowed Timing mode;
Col. (3): the net exposure time of \textit{Swift}-XRT in units of second;
Col. (4): the start time (MJD) of the \textit{Swift} observation.
Col. (5): the \textit{Swift}-XRT photon index; 
Col. (6): the error of \textit{Swift}-XRT photon index; 
Col. (7): the normalization flux in units of $\rm cm^{-2}\, s^{-1}\, keV^{-1}$; 
Col. (8): the error of normalization flux in units of $\rm cm^{-2}\, s^{-1}\, keV^{-1}$; 
Col. (9): the unabsorbed flux in 0.3$-$10 keV in units of $\rm erg \, cm^{-2} \, s^{-1}$;
Col. (10): the error of the unabsorbed flux in 0.3$-$10 keV in units of $\rm erg \, cm^{-2} \, s^{-1}$;
Col. (11): the Cash statistic;
Col. (12): the degrees of freedom. 
}
\end{deluxetable}

The \textit{Swift}-UVOT aperture photometric data after the Galactic extinction correction are shown in Tab. \ref{tab:UVOT}. 

\begin{deluxetable}{ccccccccccccccccccc}
\tabletypesize{\tiny}
\tablewidth{0pt} 
\tablecaption{\textit{Swift}-UVOT aperture photometric data \label{tab:UVOT}}
\rotate 
\tablehead{   
          & \multicolumn{3}{c}{$u$ band} & \multicolumn{3}{c}{$b$ band} & \multicolumn{3}{c}{$v$ band} & \multicolumn{3}{c}{$uvw1$ band} & \multicolumn{3}{c}{$uvw2$ band} & \multicolumn{3}{c}{$uvm2$ band} \\ 
    \cmidrule(lr){2-4} \cmidrule(lr){5-7} \cmidrule(lr){8-10} \cmidrule(lr){11-13} \cmidrule(lr){14-16} \cmidrule(lr){17-19}
    ObsID & MJD   & $F_{\lambda}$ & $F_{\lambda, \, \rm err}$ & MJD   & $F_{\lambda}$ & $F_{\lambda, \, \rm err}$ & MJD   & $F_{\lambda}$ & $F_{\lambda, \, \rm err}$ & MJD   & $F_{\lambda}$ & $F_{\lambda, \, \rm err}$ & MJD   & $F_{\lambda}$ & $F_{\lambda, \, \rm err}$ & MJD   & $F_{\lambda}$ & $F_{\lambda, \, \rm err}$
}
\colnumbers 
\startdata
    00035009143 & 57011.04 & 1.12E-14 & 4.18E-16 & 57011.04 & 1.00E-14 & 3.23E-16 & 57011.04 & 8.32E-15 & 2.87E-16 & 57011.04 & 1.30E-14 & 6.43E-16 & 57011.04 & 1.57E-14 & 6.72E-16 & 57011.05 & 1.49E-14 & 6.08E-16 \\
    00035009144 & 57019.36 & 1.24E-14 & 4.51E-16 & 57019.36 & 1.07E-14 & 3.36E-16 & 57019.36 & 8.82E-15 & 2.90E-16 & 57019.36 & 1.33E-14 & 6.51E-16 & 57019.36 & 1.61E-14 & 6.84E-16 & 57019.37 & 1.56E-14 & 6.19E-16 \\
    00035009145 & 57023.22 & 1.04E-14 & 3.84E-16 & 57023.22 & 8.71E-15 & 2.79E-16 & 57023.22 & 7.52E-15 & 2.56E-16 & 57023.22 & 1.13E-14 & 5.58E-16 & 57023.22 & 1.41E-14 & 6.05E-16 & 57023.23 & 1.37E-14 & 5.50E-16 \\
    00035009146 & 57029.01 & 1.30E-14 & 4.83E-16 & 57029.01 & 1.16E-14 & 3.75E-16 & 57029.01 & 9.70E-15 & 3.33E-16 & 57029.01 & 1.42E-14 & 7.03E-16 & 57029.01 & 1.79E-14 & 7.69E-16 & 57029.01 & 1.77E-14 & 7.10E-16 \\
    00035009147 & 57041.09 & 6.36E-14 & 2.35E-15 & 57041.09 & 5.42E-14 & 1.73E-15 & 57041.10 & 5.05E-14 & 1.40E-15 & 57041.09 & 8.53E-14 & 4.00E-15 & 57041.10 & 1.09E-13 & 4.46E-15 & 57041.10 & 1.08E-13 & 3.94E-15 \\
    00035009148 & 57042.75 & 5.32E-14 & 1.89E-15 & 57042.75 & 4.46E-14 & 1.35E-15 & 57042.76 & 3.93E-14 & 1.07E-15 & 57042.75 & 6.22E-14 & 2.91E-15 & 57042.76 & 7.64E-14 & 3.13E-15 & 57042.76 & 7.60E-14 & 2.77E-15 \\
    00035009149 & --    & --    & --    & --    & --    & --    & --    & --    & --    & --    & --    & --    & --    & --    & --    & --    & --    & -- \\
    00035009152 & 57044.08 & 4.95E-14 & 1.77E-15 & 57044.08 & 4.05E-14 & 1.23E-15 & 57044.09 & 3.60E-14 & 9.96E-16 & 57044.08 & 5.96E-14 & 2.80E-15 & 57044.09 & 7.40E-14 & 3.04E-15 & 57044.09 & 7.46E-14 & 2.72E-15 \\
    00035009153 & 57044.29 & 4.55E-14 & 1.50E-15 & --    & --    & --    & --    & --    & --    & --    & --    & --    & --    & --    & --    & --    & --    & -- \\
          & 57044.36 & 4.51E-14 & 1.47E-15 & --    & --    & --    & --    & --    & --    & --    & --    & --    & --    & --    & --    & --    & --    & -- \\
          & 57044.42 & 4.61E-14 & 1.50E-15 & --    & --    & --    & --    & --    & --    & --    & --    & --    & --    & --    & --    & --    & --    & -- \\
          & 57044.48 & 4.95E-14 & 1.63E-15 & --    & --    & --    & --    & --    & --    & --    & --    & --    & --    & --    & --    & --    & --    & -- \\
          & 57044.55 & 5.18E-14 & 1.68E-15 & --    & --    & --    & --    & --    & --    & --    & --    & --    & --    & --    & --    & --    & --    & -- \\
          & 57044.62 & 5.33E-14 & 1.74E-15 & --    & --    & --    & --    & --    & --    & --    & --    & --    & --    & --    & --    & --    & --    & -- \\
          & 57044.68 & 5.29E-14 & 1.72E-15 & --    & --    & --    & --    & --    & --    & --    & --    & --    & --    & --    & --    & --    & --    & -- \\
          & 57044.81 & 5.61E-14 & 1.84E-15 & --    & --    & --    & --    & --    & --    & --    & --    & --    & --    & --    & --    & --    & --    & -- \\
          & 57044.95 & 6.02E-14 & 1.97E-15 & --    & --    & --    & --    & --    & --    & --    & --    & --    & --    & --    & --    & --    & --    & -- \\
    00035009154 & 57045.01 & 5.76E-14 & 2.05E-15 & 57045.01 & 4.89E-14 & 1.48E-15 & 57045.02 & 4.43E-14 & 1.19E-15 & 57045.01 & 7.69E-14 & 3.59E-15 & 57045.02 & 1.00E-13 & 4.09E-15 & 57045.02 & 9.95E-14 & 3.60E-15 \\
    00035009156 & --    & --    & --    & --    & --    & --    & --    & --    & --    & --    & --    & --    & 57045.15 & 1.03E-13 & 4.18E-15 & 57046.15 & 1.28E-13 & 4.56E-15 \\
          & --    & --    & --    & --    & --    & --    & --    & --    & --    & --    & --    & --    & 57045.28 & 1.05E-13 & 4.27E-15 & 57046.22 & 1.19E-13 & 4.25E-15 \\
          & --    & --    & --    & --    & --    & --    & --    & --    & --    & --    & --    & --    & 57045.41 & 1.16E-13 & 4.71E-15 & 57046.28 & 1.10E-13 & 3.90E-15 \\
          & --    & --    & --    & --    & --    & --    & --    & --    & --    & --    & --    & --    & 57045.48 & 1.18E-13 & 4.76E-15 & 57046.40 & 1.19E-13 & 4.32E-15 \\
          & --    & --    & --    & --    & --    & --    & --    & --    & --    & --    & --    & --    & 57045.68 & 1.24E-13 & 5.03E-15 & 57046.48 & 1.14E-13 & 4.03E-15 \\
    00035009157 & 57047.14 & 7.07E-14 & 2.79E-15 & 57047.14 & 5.96E-14 & 2.05E-15 & 57047.14 & 5.67E-14 & 1.65E-15 & 57047.14 & 9.90E-14 & 4.68E-15 & 57047.48 & 1.13E-13 & 4.62E-15 & 57047.48 & 1.10E-13 & 3.98E-15 \\
          & 57047.48 & 6.44E-14 & 2.40E-15 & 57047.48 & 5.44E-14 & 1.75E-15 & 57047.48 & 4.84E-14 & 1.36E-15 & --    & --    & --    & --    & --    & --    & --    & --    & -- \\
    00035009158 & --    & --    & --    & --    & --    & --    & --    & --    & --    & 57047.22 & 1.10E-13 & 5.07E-15 & --    & --    & --    & --    & --    & -- \\
          & --    & --    & --    & --    & --    & --    & --    & --    & --    & 57047.42 & 8.51E-14 & 3.91E-15 & --    & --    & --    & --    & --    & -- \\
          & --    & --    & --    & --    & --    & --    & --    & --    & --    & 57047.47 & 8.72E-14 & 4.10E-15 & --    & --    & --    & --    & --    & -- \\
          & --    & --    & --    & --    & --    & --    & --    & --    & --    & 57047.61 & 9.68E-14 & 4.45E-15 & --    & --    & --    & --    & --    & -- \\
          & --    & --    & --    & --    & --    & --    & --    & --    & --    & 57047.67 & 1.00E-13 & 4.61E-15 & --    & --    & --    & --    & --    & -- \\
          & --    & --    & --    & --    & --    & --    & --    & --    & --    & 57047.80 & 9.43E-14 & 4.35E-15 & --    & --    & --    & --    & --    & -- \\
          & --    & --    & --    & --    & --    & --    & --    & --    & --    & 57047.87 & 8.45E-14 & 3.88E-15 & --    & --    & --    & --    & --    & -- \\
          & --    & --    & --    & --    & --    & --    & --    & --    & --    & 57047.93 & 8.22E-14 & 3.79E-15 & --    & --    & --    & --    & --    & -- \\
    00035009159 & 57048.73 & 6.38E-14 & 2.36E-15 & 57048.73 & 5.19E-14 & 1.64E-15 & 57048.73 & 4.70E-14 & 1.30E-15 & 57048.73 & 8.48E-14 & 3.98E-15 & 57048.73 & 1.09E-13 & 4.47E-15 & 57048.74 & 1.08E-13 & 3.84E-15 \\
    00035009160 & 57048.87 & 6.49E-14 & 2.11E-15 & --    & --    & --    & --    & --    & --    & --    & --    & --    & --    & --    & --    & --    & --    & -- \\
    00035009161 & --    & --    & --    & --    & --    & --    & --    & --    & --    & --    & --    & --    & --    & --    & --    & --    & --    & -- \\
    00035009162 & 57050.01 & 4.95E-14 & 1.92E-15 & 57050.01 & 4.22E-14 & 1.44E-15 & 57050.01 & 3.65E-14 & 1.14E-15 & --    & --    & --    & --    & --    & --    & --    & --    & -- \\
          & 57050.14 & 4.41E-14 & 1.70E-15 & 57050.14 & 3.58E-14 & 1.20E-15 & 57050.15 & 3.14E-14 & 9.95E-16 & --    & --    & --    & --    & --    & --    & --    & --    & -- \\
    00035009167 & 57051.26 & 3.56E-14 & 1.36E-15 & 57051.32 & 3.05E-14 & 9.21E-16 & 57051.33 & 2.63E-14 & 7.49E-16 & 57051.26 & 4.23E-14 & 2.00E-15 & 57051.33 & 5.34E-14 & 2.20E-15 & 57051.33 & 5.21E-14 & 1.86E-15 \\
          & 57051.32 & 3.56E-14 & 1.26E-15 & 57051.39 & 2.94E-14 & 8.88E-16 & 57051.40 & 2.47E-14 & 7.06E-16 & 57051.32 & 4.28E-14 & 2.02E-15 & 57051.39 & 5.12E-14 & 2.11E-15 & 57051.40 & 5.45E-14 & 6.72E-15 \\
          & 57051.39 & 3.52E-14 & 1.24E-15 & 57051.46 & 2.75E-14 & 8.31E-16 & --    & --    & --    & 57051.39 & 4.11E-14 & 1.94E-15 & 57051.46 & 4.82E-14 & 2.01E-15 & --    & --    & -- \\
          & 57051.46 & 3.31E-14 & 1.17E-15 & --    & --    & --    & --    & --    & --    & 57051.45 & 3.82E-14 & 1.81E-15 & --    & --    & --    & --    & --    & -- \\
    00035009164 & 57051.67 & 3.52E-14 & 1.32E-15 & --    & --    & --    & --    & --    & --    & 57051.66 & 4.13E-14 & 1.95E-15 & --    & --    & --    & --    & --    & -- \\
    00035009168 & 57051.93 & 3.36E-14 & 1.19E-15 & 57051.86 & 2.83E-14 & 8.54E-16 & 57051.86 & 2.40E-14 & 6.87E-16 & 57051.92 & 3.78E-14 & 1.79E-15 & 57051.93 & 4.65E-14 & 1.92E-15 & --    & --    & -- \\
          & 57051.95 & 3.30E-14 & 1.17E-15 & 57051.93 & 2.77E-14 & 8.35E-16 & 57051.93 & 2.37E-14 & 6.97E-16 & 57051.95 & 3.83E-14 & 1.81E-15 & --    & --    & --    & --    & --    & -- \\
          & 57051.99 & 3.06E-14 & 1.08E-15 & 57051.99 & 2.66E-14 & 8.04E-16 & --    & --    & --    & --    & --    & --    & --    & --    & --    & --    & --    & -- \\
    00035009169 & 57052.27 & 3.04E-14 & 9.93E-16 & 57052.79 & 3.01E-14 & 8.60E-16 & 57052.80 & 2.65E-14 & 6.96E-16 & --    & --    & --    & --    & --    & --    & --    & --    & -- \\
          & 57052.79 & 3.47E-14 & 1.18E-15 & --    & --    & --    & --    & --    & --    & --    & --    & --    & --    & --    & --    & --    & --    & -- \\
          & 57052.86 & 3.54E-14 & 1.15E-15 & --    & --    & --    & --    & --    & --    & --    & --    & --    & --    & --    & --    & --    & --    & -- \\
          & 57052.99 & 3.28E-14 & 1.08E-15 & --    & --    & --    & --    & --    & --    & --    & --    & --    & --    & --    & --    & --    & --    & -- \\
    00035009170 & 57053.00 & 3.20E-14 & 1.04E-15 & 57053.73 & 2.22E-14 & 6.58E-16 & 57053.73 & 1.88E-14 & 5.34E-16 & --    & --    & --    & 57053.07 & 4.79E-14 & 1.95E-15 & --    & --    & -- \\
          & 57053.73 & 2.62E-14 & 9.14E-16 & --    & --    & --    & --    & --    & --    & --    & --    & --    & 57053.14 & 4.21E-14 & 1.71E-15 & --    & --    & -- \\
          & --    & --    & --    & --    & --    & --    & --    & --    & --    & --    & --    & --    & 57053.20 & 4.19E-14 & 1.70E-15 & --    & --    & -- \\
          & --    & --    & --    & --    & --    & --    & --    & --    & --    & --    & --    & --    & 57053.54 & 4.15E-14 & 1.70E-15 & --    & --    & -- \\
    00035009171 & 57054.65 & 2.78E-14 & 9.48E-16 & 57054.65 & 2.34E-14 & 6.72E-16 & 57054.66 & 1.94E-14 & 5.24E-16 & 57054.65 & 3.29E-14 & 1.54E-15 & 57054.66 & 4.03E-14 & 1.65E-15 & 57054.59 & 4.04E-14 & 1.44E-15 \\
          & --    & --    & --    & --    & --    & --    & --    & --    & --    & --    & --    & --    & --    & --    & --    & 57054.66 & 3.92E-14 & 1.44E-15 \\
    00035009172 & 57055.45 & 3.59E-14 & 1.27E-15 & 57055.45 & 2.93E-14 & 8.80E-16 & 57055.46 & 2.44E-14 & 6.94E-16 & 57055.45 & 4.15E-14 & 1.96E-15 & 57055.45 & 5.20E-14 & 2.14E-15 & 57055.46 & 5.03E-14 & 1.85E-15 \\
    00035009173 & 57056.31 & 4.42E-14 & 1.58E-15 & 57056.31 & 3.54E-14 & 1.08E-15 & 57056.32 & 3.09E-14 & 8.71E-16 & 57056.31 & 5.44E-14 & 2.56E-15 & 57056.32 & 6.94E-14 & 2.85E-15 & 57056.32 & 6.80E-14 & 2.47E-15 \\
    00035009174 & 57057.64 & 4.21E-14 & 1.50E-15 & 57057.64 & 3.54E-14 & 1.08E-15 & 57057.65 & 2.90E-14 & 8.17E-16 & --    & --    & --    & 57057.64 & 6.85E-14 & 2.81E-15 & 57057.65 & 6.69E-14 & 2.39E-15 \\
    00035009175 & 57058.52 & 4.43E-14 & 1.60E-15 & 57058.52 & 3.68E-14 & 1.13E-15 & 57058.53 & 3.06E-14 & 8.75E-16 & 57058.52 & 5.50E-14 & 2.59E-15 & 57058.52 & 7.11E-14 & 2.93E-15 & 57058.53 & 6.94E-14 & 2.53E-15 \\
          & 57058.59 & 4.37E-14 & 1.54E-15 & 57058.59 & 3.52E-14 & 1.06E-15 & 57058.59 & 2.98E-14 & 8.29E-16 & 57058.59 & 5.48E-14 & 2.57E-15 & 57058.59 & 7.00E-14 & 2.87E-15 & 57058.60 & 6.79E-14 & 2.47E-15 \\
    00035009176 & --    & --    & --    & --    & --    & --    & --    & --    & --    & --    & --    & --    & --    & --    & --    & --    & --    & -- \\
    00035009177 & 57059.65 & 3.99E-14 & 1.36E-15 & 57059.65 & 3.28E-14 & 9.44E-16 & 57059.66 & 2.75E-14 & 7.44E-16 & 57059.05 & 5.07E-14 & 2.34E-15 & 57059.65 & 6.26E-14 & 2.55E-15 & --    & --    & -- \\
          & --    & --    & --    & --    & --    & --    & --    & --    & --    & 57059.12 & 5.29E-14 & 2.43E-15 & --    & --    & --    & --    & --    & -- \\
          & --    & --    & --    & --    & --    & --    & --    & --    & --    & 57059.19 & 5.23E-14 & 2.40E-15 & --    & --    & --    & --    & --    & -- \\
          & --    & --    & --    & --    & --    & --    & --    & --    & --    & 57059.31 & 5.20E-14 & 2.39E-15 & --    & --    & --    & --    & --    & -- \\
          & --    & --    & --    & --    & --    & --    & --    & --    & --    & 57059.52 & 5.27E-14 & 2.43E-15 & --    & --    & --    & --    & --    & -- \\
          & --    & --    & --    & --    & --    & --    & --    & --    & --    & 57059.58 & 5.24E-14 & 2.41E-15 & --    & --    & --    & --    & --    & -- \\
          & --    & --    & --    & --    & --    & --    & --    & --    & --    & 57059.64 & 4.90E-14 & 2.28E-15 & --    & --    & --    & --    & --    & -- \\
          & --    & --    & --    & --    & --    & --    & --    & --    & --    & 57059.71 & 4.55E-14 & 2.10E-15 & --    & --    & --    & --    & --    & -- \\
          & --    & --    & --    & --    & --    & --    & --    & --    & --    & 57059.85 & 4.60E-14 & 2.11E-15 & --    & --    & --    & --    & --    & -- \\
          & --    & --    & --    & --    & --    & --    & --    & --    & --    & 57059.92 & 4.54E-14 & 2.09E-15 & --    & --    & --    & --    & --    & -- \\
          & --    & --    & --    & --    & --    & --    & --    & --    & --    & 57059.98 & 4.54E-14 & 2.09E-15 & --    & --    & --    & --    & --    & -- \\
    00035009178 & 57060.05 & 3.65E-14 & 1.20E-15 & 57060.98 & 2.87E-14 & 8.23E-16 & 57060.99 & 2.34E-14 & 6.24E-16 & 57060.98 & 4.22E-14 & 1.97E-15 & 57060.98 & 5.31E-14 & 2.17E-15 & 57060.99 & 5.06E-14 & 1.84E-15 \\
          & 57060.12 & 3.79E-14 & 1.24E-15 & --    & --    & --    & --    & --    & --    & --    & --    & --    & --    & --    & --    & --    & --    & -- \\
          & 57060.19 & 3.97E-14 & 1.29E-15 & --    & --    & --    & --    & --    & --    & --    & --    & --    & --    & --    & --    & --    & --    & -- \\
          & 57060.25 & 3.83E-14 & 1.24E-15 & --    & --    & --    & --    & --    & --    & --    & --    & --    & --    & --    & --    & --    & --    & -- \\
          & 57060.32 & 3.83E-14 & 1.24E-15 & --    & --    & --    & --    & --    & --    & --    & --    & --    & --    & --    & --    & --    & --    & -- \\
          & 57060.39 & 3.85E-14 & 1.25E-15 & --    & --    & --    & --    & --    & --    & --    & --    & --    & --    & --    & --    & --    & --    & -- \\
          & 57060.52 & 3.63E-14 & 1.18E-15 & --    & --    & --    & --    & --    & --    & --    & --    & --    & --    & --    & --    & --    & --    & -- \\
          & 57060.58 & 3.51E-14 & 1.14E-15 & --    & --    & --    & --    & --    & --    & --    & --    & --    & --    & --    & --    & --    & --    & -- \\
          & 57060.72 & 3.47E-14 & 1.13E-15 & --    & --    & --    & --    & --    & --    & --    & --    & --    & --    & --    & --    & --    & --    & -- \\
          & 57060.79 & 3.48E-14 & 1.15E-15 & --    & --    & --    & --    & --    & --    & --    & --    & --    & --    & --    & --    & --    & --    & -- \\
          & 57060.86 & 3.41E-14 & 1.12E-15 & --    & --    & --    & --    & --    & --    & --    & --    & --    & --    & --    & --    & --    & --    & -- \\
          & 57060.91 & 3.41E-14 & 1.10E-15 & --    & --    & --    & --    & --    & --    & --    & --    & --    & --    & --    & --    & --    & --    & -- \\
          & 57060.98 & 3.43E-14 & 1.17E-15 & --    & --    & --    & --    & --    & --    & --    & --    & --    & --    & --    & --    & --    & --    & -- \\
    00035009179 & 57061.24 & 3.21E-14 & 1.09E-15 & 57061.24 & 2.72E-14 & 7.78E-16 & 57061.25 & 2.24E-14 & 5.98E-16 & 57061.23 & 3.95E-14 & 1.84E-15 & 57061.05 & 5.53E-14 & 2.24E-15 & 57061.25 & 4.83E-14 & 1.75E-15 \\
          & --    & --    & --    & --    & --    & --    & --    & --    & --    & --    & --    & --    & 57061.12 & 5.29E-14 & 2.14E-15 & --    & --    & -- \\
          & --    & --    & --    & --    & --    & --    & --    & --    & --    & --    & --    & --    & 57061.24 & 4.97E-14 & 2.03E-15 & --    & --    & -- \\
          & --    & --    & --    & --    & --    & --    & --    & --    & --    & --    & --    & --    & 57061.52 & 4.00E-14 & 1.62E-15 & --    & --    & -- \\
          & --    & --    & --    & --    & --    & --    & --    & --    & --    & --    & --    & --    & 57061.57 & 4.20E-14 & 1.73E-15 & --    & --    & -- \\
          & --    & --    & --    & --    & --    & --    & --    & --    & --    & --    & --    & --    & 57061.78 & 4.77E-14 & 1.93E-15 & --    & --    & -- \\
    00035009180 & --    & --    & --    & --    & --    & --    & --    & --    & --    & --    & --    & --    & --    & --    & --    & 57062.06 & 3.84E-14 & 1.44E-15 \\
          & --    & --    & --    & --    & --    & --    & --    & --    & --    & --    & --    & --    & --    & --    & --    & 57062.17 & 3.90E-14 & 1.39E-15 \\
          & --    & --    & --    & --    & --    & --    & --    & --    & --    & --    & --    & --    & --    & --    & --    & 57062.24 & 4.26E-14 & 1.52E-15 \\
          & --    & --    & --    & --    & --    & --    & --    & --    & --    & --    & --    & --    & --    & --    & --    & 57062.31 & 4.11E-14 & 1.47E-15 \\
          & --    & --    & --    & --    & --    & --    & --    & --    & --    & --    & --    & --    & --    & --    & --    & 57062.37 & 4.03E-14 & 1.48E-15 \\
          & --    & --    & --    & --    & --    & --    & --    & --    & --    & --    & --    & --    & --    & --    & --    & 57062.44 & 3.91E-14 & 1.39E-15 \\
          & --    & --    & --    & --    & --    & --    & --    & --    & --    & --    & --    & --    & --    & --    & --    & 57062.50 & 3.84E-14 & 1.37E-15 \\
          & --    & --    & --    & --    & --    & --    & --    & --    & --    & --    & --    & --    & --    & --    & --    & 57062.57 & 3.85E-14 & 1.37E-15 \\
          & --    & --    & --    & --    & --    & --    & --    & --    & --    & --    & --    & --    & --    & --    & --    & 57062.64 & 3.75E-14 & 1.34E-15 \\
          & --    & --    & --    & --    & --    & --    & --    & --    & --    & --    & --    & --    & --    & --    & --    & 57062.70 & 3.71E-14 & 1.32E-15 \\
    00035009181 & 57063.96 & 2.46E-14 & 8.60E-16 & 57063.96 & 2.08E-14 & 6.22E-16 & 57063.97 & 1.75E-14 & 5.07E-16 & 57063.96 & 2.94E-14 & 1.39E-15 & 57063.97 & 3.74E-14 & 1.54E-15 & 57063.97 & 3.68E-14 & 1.38E-15 \\
    00035009182 & 57064.30 & 2.59E-14 & 9.17E-16 & 57064.30 & 2.15E-14 & 6.54E-16 & 57064.31 & 1.80E-14 & 5.34E-16 & 57064.30 & 3.07E-14 & 1.46E-15 & 57064.31 & 3.92E-14 & 1.63E-15 & 57064.31 & 3.70E-14 & 1.39E-15 \\
    00035009184 & 57066.03 & 3.10E-14 & 1.09E-15 & 57066.04 & 2.44E-14 & 7.29E-16 & --    & --    & --    & 57066.03 & 3.58E-14 & 1.69E-15 & 57066.04 & 4.68E-14 & 1.94E-15 & 57066.10 & 4.39E-14 & 1.57E-15 \\
    00035009185 & --    & --    & --    & --    & --    & --    & --    & --    & --    & --    & --    & --    & --    & --    & --    & 57066.11 & 4.36E-14 & 1.56E-15 \\
          & --    & --    & --    & --    & --    & --    & --    & --    & --    & --    & --    & --    & --    & --    & --    & 57066.17 & 4.42E-14 & 1.58E-15 \\
          & --    & --    & --    & --    & --    & --    & --    & --    & --    & --    & --    & --    & --    & --    & --    & 57066.23 & 4.47E-14 & 1.60E-15 \\
          & --    & --    & --    & --    & --    & --    & --    & --    & --    & --    & --    & --    & --    & --    & --    & 57066.24 & 4.40E-14 & 1.57E-15 \\
          & --    & --    & --    & --    & --    & --    & --    & --    & --    & --    & --    & --    & --    & --    & --    & 57066.30 & 4.22E-14 & 1.51E-15 \\
          & --    & --    & --    & --    & --    & --    & --    & --    & --    & --    & --    & --    & --    & --    & --    & 57066.31 & 4.27E-14 & 1.53E-15 \\
          & --    & --    & --    & --    & --    & --    & --    & --    & --    & --    & --    & --    & --    & --    & --    & 57066.36 & 4.34E-14 & 1.55E-15 \\
    00035009186 & --    & --    & --    & --    & --    & --    & --    & --    & --    & --    & --    & --    & --    & --    & --    & 57066.37 & 4.40E-14 & 1.58E-15 \\
    00035009190 & 57067.91 & 3.06E-14 & 1.08E-15 & 57067.91 & 2.57E-14 & 7.75E-16 & 57067.97 & 2.09E-14 & 6.08E-16 & 57067.96 & 3.75E-14 & 1.77E-15 & 57067.97 & 4.91E-14 & 2.03E-15 & --    & --    & -- \\
          & 57067.97 & 3.12E-14 & 1.10E-15 & 57067.97 & 2.57E-14 & 7.75E-16 & --    & --    & --    & --    & --    & --    & --    & --    & --    & --    & --    & -- \\
    00035009187 & 57068.30 & 2.77E-14 & 9.63E-16 & --    & --    & --    & --    & --    & --    & 57068.30 & 3.36E-14 & 1.58E-15 & 57068.30 & 4.24E-14 & 1.75E-15 & 57068.29 & 4.17E-14 & 1.54E-15 \\
    00035009188 & --    & --    & --    & --    & --    & --    & --    & --    & --    & 57068.82 & 2.59E-14 & 1.24E-15 & 57068.83 & 3.22E-14 & 1.34E-15 & 57068.82 & 3.11E-14 & 1.18E-15 \\
    00035009189 & 57069.96 & 2.04E-14 & 7.10E-16 & --    & --    & --    & --    & --    & --    & 57069.96 & 2.42E-14 & 1.14E-15 & 57069.96 & 2.98E-14 & 1.24E-15 & --    & --    & -- \\
    00035009192 & 57070.23 & 2.02E-14 & 7.01E-16 & --    & --    & --    & --    & --    & --    & 57070.23 & 2.32E-14 & 1.10E-15 & 57070.24 & 2.87E-14 & 1.19E-15 & 57070.23 & 2.86E-14 & 1.07E-15 \\
    00035009193 & 57070.76 & 1.77E-14 & 6.16E-16 & --    & --    & --    & --    & --    & --    & 57070.76 & 2.00E-14 & 9.51E-16 & 57070.76 & 2.51E-14 & 1.04E-15 & --    & --    & -- \\
    00035009194 & 57071.09 & 2.01E-14 & 7.11E-16 & --    & --    & --    & --    & --    & --    & 57071.09 & 2.36E-14 & 1.12E-15 & 57071.10 & 3.06E-14 & 1.27E-15 & 57071.09 & 2.95E-14 & 1.12E-15 \\
    00035009195 & 57071.69 & 2.17E-14 & 7.58E-16 & --    & --    & --    & --    & --    & --    & 57071.69 & 2.44E-14 & 1.15E-15 & 57071.70 & 3.14E-14 & 1.30E-15 & 57071.69 & 3.00E-14 & 1.12E-15 \\
    00035009196 & 57072.02 & 2.17E-14 & 7.61E-16 & --    & --    & --    & --    & --    & --    & 57072.02 & 2.52E-14 & 1.20E-15 & 57072.02 & 3.15E-14 & 1.30E-15 & 57072.75 & 3.80E-14 & 1.50E-15 \\
    00035009197 & 57072.75 & 2.69E-14 & 1.02E-15 & --    & --    & --    & --    & --    & --    & 57072.75 & 3.06E-14 & 1.49E-15 & 57072.75 & 3.95E-14 & 1.66E-15 & --    & --    & -- \\
    00035009198 & 57073.02 & 2.27E-14 & 7.93E-16 & --    & --    & --    & --    & --    & --    & 57073.01 & 2.69E-14 & 1.27E-15 & 57073.02 & 3.43E-14 & 1.41E-15 & 57073.01 & 3.33E-14 & 1.24E-15 \\
    00035009199 & 57073.61 & 2.14E-14 & 7.50E-16 & --    & --    & --    & --    & --    & --    & 57073.61 & 2.39E-14 & 1.13E-15 & 57073.62 & 3.13E-14 & 1.29E-15 & 57073.61 & 3.01E-14 & 1.13E-15 \\
    00035009200 & 57074.15 & 1.65E-14 & 5.84E-16 & --    & --    & --    & --    & --    & --    & 57074.15 & 1.93E-14 & 9.20E-16 & 57074.15 & 2.38E-14 & 9.91E-16 & 57074.14 & 2.35E-14 & 8.94E-16 \\
    00035009201 & 57074.61 & 1.46E-14 & 5.19E-16 & --    & --    & --    & --    & --    & --    & 57074.61 & 1.71E-14 & 8.21E-16 & 57074.62 & 2.17E-14 & 9.05E-16 & 57074.61 & 2.12E-14 & 8.09E-16 \\
\enddata
\tablecomments{
These optical fluxes are corrected for the Galactic extinction. 
The units of $F_{\lambda}$ is erg cm$^{-2}$ s$^{-1}$ $\mathring{A}^{-1}$. 
The symbol ``--" represents a null value. 
}
\end{deluxetable}

\bibliography{Ref}{}
\bibliographystyle{aasjournal}

\end{document}